\documentclass[twocolumn,tighten]{aastex63}
\usepackage{amssymb}
\usepackage{array,multirow}
\usepackage{color}

\newcommand{\gyr}{{\rm{Gyr}}}

\newcommand{\msun}{{\rm{M_\odot}}}

\newcommand{\pc}{{\rm{pc}}}
\newcommand{\henon}{{H\'enon}}
\newcommand{\metal}{{\rm{Z}}}
\newcommand{\kpc}{{\rm{kpc}}}
\newcommand{\rcobs}{{r_{c,\rm{obs}}}}
\newcommand{\rhl}{{r_{\rm{hl}}}}
\newcommand{\Sigmacobs}{{\Sigma_{c,\rm{obs}}}}
\newcommand{\pop}{{\tt{Pop}}}
\newcommand{\popone}{{\tt{Pop1}}}
\newcommand{\poptwo}{{\tt{Pop2}}}
\newcommand{\popthree}{{\tt{Pop3}}}
\newcommand{\popfour}{{\tt{Pop4}}}
\newcommand{\Lto}{{L_{\rm{TO}}}}
\newcommand{\Deltarfifty}{{\Delta_{r50}}}
\newcommand{\Deltaa}{{\Delta_A}}
\newcommand{\nbh}{{N_{\rm{BH}}}}
\newcommand{\mbh}{{M_{\rm{BH}}}}
\newcommand{\ncluster}{{N_{\rm{cluster}}}}
\newcommand{\mcluster}{{M_{\rm{cluster}}}}
\newcommand{\lcluster}{{L_{\rm{cluster}}}}
\newcommand{\rlim}{{r_{\rm{lim}}}}

\newcommand{\pmsigma}{\pm 1\sigma}

\newlength{\mysize}

\shorttitle{Constraining BH Populations in 50 MWGCs}
\shortauthors{Weatherford, et al}

\begin{document}
\title{A Dynamical Survey of Stellar-Mass Black Holes in 50 Milky Way Globular Clusters}

\author[0000-0002-9660-9085]{Newlin C. Weatherford}
\affil{Department of Physics \& Astronomy, Northwestern University, Evanston, IL 60208, USA}
\affil{Center for Interdisciplinary Exploration \& Research in Astrophysics (CIERA), Northwestern University, Evanston, IL 60208, USA}
\email{newlinweatherford2017@u.northwestern.edu}

\author[0000-0002-3680-2684]{Sourav Chatterjee}
\affil{Tata Institute of Fundamental Research, Homi Bhabha Road, Mumbai 400005, India}
\email{souravchatterjee.tifr@gmail.com}

\author[0000-0002-4086-3180]{Kyle Kremer}
\affil{Department of Physics \& Astronomy, Northwestern University, Evanston, IL 60208, USA}
\affil{Center for Interdisciplinary Exploration \& Research in Astrophysics (CIERA), Northwestern University, Evanston, IL 60208, USA}

\author[0000-0002-7132-418X]{Frederic A. Rasio}
\affil{Department of Physics \& Astronomy, Northwestern University, Evanston, IL 60208, USA}
\affil{Center for Interdisciplinary Exploration \& Research in Astrophysics (CIERA), Northwestern University, Evanston, IL 60208, USA}

\vspace{0.3cm}
\begin{abstract}
Recent numerical simulations of globular clusters (GCs) have shown that stellar-mass black holes (BHs) play a fundamental role in driving cluster evolution and shaping their present-day structure. Rapidly mass-segregating to the center of GCs, BHs act as a dynamical energy source via repeated super-elastic scattering, delaying onset of core collapse and limiting mass segregation for visible stars. While recent discoveries of BH candidates in Galactic and extragalactic GCs have further piqued interest in BH-mediated cluster dynamics, numerical models show that even if significant BH populations remain in today's GCs, they are not typically in directly detectable configurations. We demonstrated in \citet{Weatherford2018} that an anti-correlation between a suitable measure of mass segregation ($\Delta$) in observable stellar populations and the number of retained BHs in GC models can be applied to indirectly probe BH populations in real GCs. Here, we estimate the number and total mass of BHs in 50 Milky Way GCs from the ACS Globular Cluster Survey. For each GC, $\Delta$ is measured between observed main sequence populations and fed into correlations between $\Delta$ and BH retention found in our \texttt{CMC} Cluster Catalog's models. We demonstrate the range in measured $\Delta$ from our models matches that for observed GCs to a remarkable degree. Our results constitute the largest sample of GCs for which BH populations have been predicted to-date using a self-consistent and robust statistical approach. We identify NGCs 2808, 5927, 5986, 6101, and 6205 to retain especially large BH populations, each with total BH mass exceeding $10^3\,\msun$.
\vspace{0.7cm}
\end{abstract}


\section{Introduction} \label{intro}
Our understanding of stellar-mass black hole (BH) populations in globular clusters (GCs) has rapidly improved since the turn of the century. To date, five BH candidates have been detected in Milky Way GCs via X-ray and radio observations: 
two in M\,22 \citep{Strader2012}, plus one each in M\,62 \citep{Chomiuk2013}, 47\,Tuc \citep{Miller-Jones2015,Bahramian2017}, and M\,10 \citep{Shishkovsky2018}. More recently, three BHs in detached binaries have been reported in NGC 3201, the first to be identified using radial velocity measurements \citep{Giesers2018,Giesers2019}. Additional candidates have been spotted in extragalactic GCs \citep[e.g.,][]{Maccarone2007,Irwin2010}.
The lack of any particular pattern in the GCs hosting BH candidates suggests that perhaps most GCs in the Milky Way (MWGCs) retain BH populations to present.

Such observational evidence complements a number of recent computational simulations which show that realistic clusters can retain up to thousands of BHs late in their lifetimes \citep[e.g.,][]{Morscher2015}.
It is now clear that BH populations play a significant role in driving long-term cluster evolution and shaping the present-day structure of GCs \citep{Merritt2004,Mackey2007,Mackey2008,BreenHeggie2013,Peuten2016,Wang2016,Chatterjee2017a,Chatterjee2017b,Weatherford2018,ArcaSedda2018,Kremer2018b,Zochi2019,Kremer2019a,Antonini2020,Kremer2020}.

The dynamical importance of BHs in GCs is reflected in their ability to explain the bimodal distribution in core radii distinguishing so-called `core-collapsed' clusters from non-core-collapsed clusters. A convincing explanation for this bimodality, specifically why most GCs are \textit{not} core-collapsed despite their short relaxation times, has challenged stellar dynamicists for decades. However, recent work by \cite{Kremer2019a,Kremer2020} has shown that cluster models naturally reproduce the range of observed cluster properties (such as core radius) when their initial size is varied within a narrow range consistent with the measured radii of young clusters in the local universe \citep{PortegiesZwart2010}. The missing piece in the explanation is simply the BHs, which guide a young cluster's evolution to manifest present-day structural features. In this picture, most clusters retain a dynamically-significant number of BHs to the present. As the BHs mass-segregate to the cluster core, they provide enough energy to passing stars in scattering interactions (via two body relaxation) to support the cluster against gravothermal collapse, at least until their ejection from the cluster \citep{Mackey2008}. For an in-depth discussion of this `BH burning' process, see \citet{BreenHeggie2013,Kremer2020}. Clusters born with high central densities rapidly extract the BH-driven dynamical energy, ejecting nearly all BHs by the present. With the ensuing reduction in dynamical energy through BH burning, the BH-poor clusters swiftly contract to the observed core-collapsed state.

Despite these advances to our understanding of BH dynamics among the cluster modeling community, observationally inferring the presence of a stellar-mass BH subsystem (BHS) in the core of a GC remains difficult. Contrary to expectations, results from $N$-body simulations suggest that the number of mass-transferring BH binaries in a GC does not correlate with the total number of BHs in the GC at the time \citep{Chatterjee2017a,Kremer2018a}. Since the majority of BH candidates in GCs come from this mass-transferring channel, the observations to-date are of little use in constraining the overall number and mass of BHs
remaining in clusters. Several groups have suggested that the existence of a BHS in a GC can be indirectly inferred from structural features, such as a large core radius and low central density \citep[e.g.,][]{Merritt2004,Hurley2007,Morscher2015,Chatterjee2017a,Askar2017a,ArcaSedda2018}. However, interpretation of such features is ambiguous; the cluster could be puffy due to BH dynamics-mediated energy production or simply because it was born puffy (equivalently, with a long initial relaxation time). Others have suggested that radial variation in the present-day stellar mass function slope may reveal the presence of a BHS \citep[e.g.,][]{Webb2016,Webb2017}. The challenge here is that obtaining enough coverage of a real GC to measure its mass function over a wide range in radial position requires consolidating observations from different space- and ground-based instruments.

Due to the above ambiguities in interpreting a GC's large-scale structural features and observational difficulties in finding its mass-function slope, we recently introduced a new approach to predict the BH content in GCs using mass segregation among visible stars from different mass ranges \citep[][W1 hereafter]{Weatherford2018}. In a journey towards energy equipartition, heavier objects in a cluster give kinetic energy to passing lighter objects through scattering interactions (two body relaxation), eventually depositing the most massive objects (the BHs) at the center, with increasingly lighter stars distributed further and further away, on average \citep[e.g.,][]{BinneyTremaine1987,HeggieHut2003}. 
The most massive stars mass-segregate closest to the central BH population, thereby undergoing closer and more frequent scattering interactions with the BHs than do less massive stars distributed further away. While BH burning drives all non-BHs away from the cluster center, the heavier objects receive proportionally more energy through this process. So, increasing the number (total mass) of BHs decreases the radial `gap' between the distributions of higher-mass and lower-mass stars. The presence of a central BH population thereby quenches mass segregation \citep[e.g.,][]{Mackey2008,Alessandrini2016}, an effect we can quantify by comparing the relative locations of stars from different mass ranges.

Low levels of mass segregation were first used to infer the existence of an intermediate-mass BH (IMBH) at the center of a GC over a decade ago \citep{Baumgardt2004,Trenti2007}.
More recently, \cite{Pasquato2016} used such a measure to place upper limits on the mass of potential IMBHs in MWGCs. \cite{Peuten2016} further suggested the lack of mass segregation between blue stragglers and stars near the main sequence turnoff in NGC 6101 may be due to an undetected BH population. W1, however, was the first study to use mass segregation to predict the number of stellar-mass BHs retained in specific MWGCs (47\,Tuc, M\,10, and M\,22). In this study, we improve upon the method first presented in W1 and apply it to predict the number ($\nbh$) and total mass ($\mbh$) of stellar-mass BH populations in 50 MWGCs from the ACS Globular Cluster Survey \citep{Sarajedini2007}.

We describe our models and how they are `observed' in \autoref{S:models}. In \autoref{S:model_results}, we define the stellar populations used to quantify mass segregation ($\Delta$), describe how we measure $\Delta$ in MWGCs from the ACS Globular Cluster Survey, and detail the steps necessary to accurately compare $\Delta$ measured in our models to $\Delta$ measured in observed clusters. We present our own present-day $\nbh$ and $\mbh$ predictions for 50 MWGCs in \autoref{S:nbh_predictions}, discuss how they support our BH burning model in \autoref{S:corecollapse}, and finally compare the predictions to previous results (most notably from the MOCCA collaboration) in \autoref{S:comparison}. Finally, in \autoref{S:summary}, we summarize all key findings and discuss a few potential wider interpretations of our results regarding primordial mass segregation and IMBHs hosted in MWGCs.

%
\section{Numerical Models} \label{S:models}
%
In this paper, we use the large grid of 148 cluster simulations presented in the \texttt{CMC} Cluster Catalog \citep{Kremer2020}, computed using the latest version of our \henon-type \citep{Henon1971a,Henon1971b} Cluster Monte Carlo code (\texttt{CMC}). \texttt{CMC} has been developed and rigorously tested over the last two decades \citep{Joshi2000, Joshi2001, Fregeau2003, Fregeau2007, Chatterjee2010, Umbreit2012, Pattabiraman2013, Chatterjee2013}. For the most recent updates and validation of \texttt{CMC} see \citet{Morscher2015, Rodriguez2016, Rodriguez2018, Kremer2020}.

As described by \citet{Kremer2020}, the grid covers roughly the full parameter space spanned by the MWGCs, with the range of variations motivated by observational constraints from high-mass young star clusters, thought to be similar in properties to GC progenitors \citep[e.g.,][]{Scheepmaker2007,Chatterjee2010}. We vary four initial parameters: the total number of particles ($N=2\times10^5$, $4\times10^5$, $8\times10^5$, and $1.6\times10^6$), the cluster virial radius ($r_v/\pc=0.5,\,1,\,2,\,4$), the metallicity ($\metal/\metal_\odot=0.01,\,0.1,\,1$), and the Galactocentric distance ($R_{\rm{gc}}/\kpc=2,\,8,\,20$) assuming a Milky Way-like galactic potential \citep[e.g.,][]{Dehnen1998}. This yields a $4\times4\times3\times3$ grid of 144 simulations. We also run four additional simulations with $N=3.2\times10^6$ particles to characterize the most massive MWGCs. For these, we fix the Galactocentric distance to $R_{\rm{gc}}/\kpc=20$ while varying metallicity ($\metal/\metal_\odot=0.01,\,1$) and virial radius ($r_v/\pc=1,\,2$). Finally, note that we exclude a handful of simulations which disrupted before reaching $13\,\gyr$ in age \citep[described in][]{Kremer2020} to ensure that our results are not affected by clusters close to disruption -- at that point, the assumption of spherical symmetry in \texttt{CMC} is incorrect. In total, we use $118$ simulations, each with a unique combination of initial properties.

In all simulations, the positions and velocities of single stars and centers of mass of binaries are drawn from a King profile with concentration $w_0=5$ \citep{King1966}. Stellar masses (primary mass in case of a binary) are drawn from the initial mass function (IMF) given in \citet{Kroupa2001} between $0.08$ and $150\,\msun$. Binaries are assigned by randomly choosing $N\times f_b$ stars independent of radial position and mass and assigning a secondary adopting a uniform mass ratio ($q$) between $0.08/m_p$ and $1$, where $m_p$ denotes the primary mass
and the binary fraction is set to $f_b = 0.05$ in all models. Binary orbital periods are drawn from a distribution flat in log scale with bounds from near contact to the hard-soft boundary. Binary eccentricities are drawn from a thermal distribution. We include all relevant physical processes, such as two body relaxation, strong binary-mediated scattering, and galactic tides using the prescriptions outlined in \citet{Kremer2020}.

Single and binary stellar evolution are followed using the \texttt{SSE} and \texttt{BSE} packages \citep{Hurley2000,Hurley2002}, updated to include our latest understanding of stellar winds \citep[e.g.,][]{Vink2001,Belczynski2010} and BH formation physics \citep[e.g.,][]{Belczynski2002,Fryer2012}. Neutron stars (NS) receive natal kicks drawn from a Maxwellian with $\sigma=265\,{\rm{km\,s^{-1}}}$. The maximum NS mass is fixed at $3\,\msun$; any remnant above this mass is considered a BH. The BH mass spectrum depends on the metallicities and pre-collapse mass \citep{Fryer2012}. BH natal kicks are based on results from \citet{Belczynski2002,Fryer2012}. Namely, a velocity is first drawn from a Maxwellian with $\sigma=265\,{\rm{km\,s^{-1}}}$, then scaled down based on the metallicity-dependent fallback of mass ejected due to supernova. These prescriptions lead to $\sim10^{-3}N$ retention of BHs immediately after they form.
By the late times of interest ($t \geq 9\,\gyr$), our simulated clusters retain a median of 3\% (0--17\%) and 2\% (0--32\%) of the total formed NSs and BHs, respectively.
More detailed descriptions and justifications are given in past work \citep[e.g.,][]{Morscher2015,Wang2016,Askar2017a}. However, note that the primary results in this work do not depend on the exact prescriptions for BH natal kicks, provided that a dynamically significant BH population remains in the cluster post-supernova. These results are expected to depend indirectly on the BH birth mass function, via modest differences it may create in the cluster's average stellar mass at late times.

\subsection{`Observing' Model Clusters} \label{S:observing_cluster_models}
\texttt{CMC} periodically outputs dynamical and stellar properties of all single and binary stars including the luminosity ($L$), temperature ($T$), and radial positions. Assuming spherical symmetry, we project the radial positions of all single and binary stars in two dimensions to create sky-projected snapshots of simulations at different times. In line with the typical age range of MWGCs, we use all snapshots ($7,355$ total or $\sim60$ per simulation) corresponding to ages between $9$ and $13\,\gyr$.

For each single star we calculate the temperature $T$ from the luminosity $L$ and the stellar radius $R$ (given by \texttt{BSE}) assuming a blackbody. We treat binaries as unresolved sources, assigning the combined luminosity $L=L_1+L_2$ and an effective temperature given by the $L$-weighted mean (Eq. 1; W1).

To account for statistical fluctuations, we project each snapshot in two dimensions assuming ten random viewing angles. For each 2D projection, we then
calculate the core radius ($\rcobs$) and central surface luminosity density ($\Sigmacobs$) by fitting an analytic approximation of the King model \citep[Eq. 18;][]{King1962} to the cumulative luminosity profile \citep[e.g.,][]{Chatterjee2017a}. We also calculate the half-light radius ($\rhl$) as the sky-projected distance from the center within which half of the total cluster light is emitted.

%
\section{Mass Segregation in Models and Observed Clusters} \label{S:model_results}
%

\begin{figure}
\epsscale{1.175}
\plotone{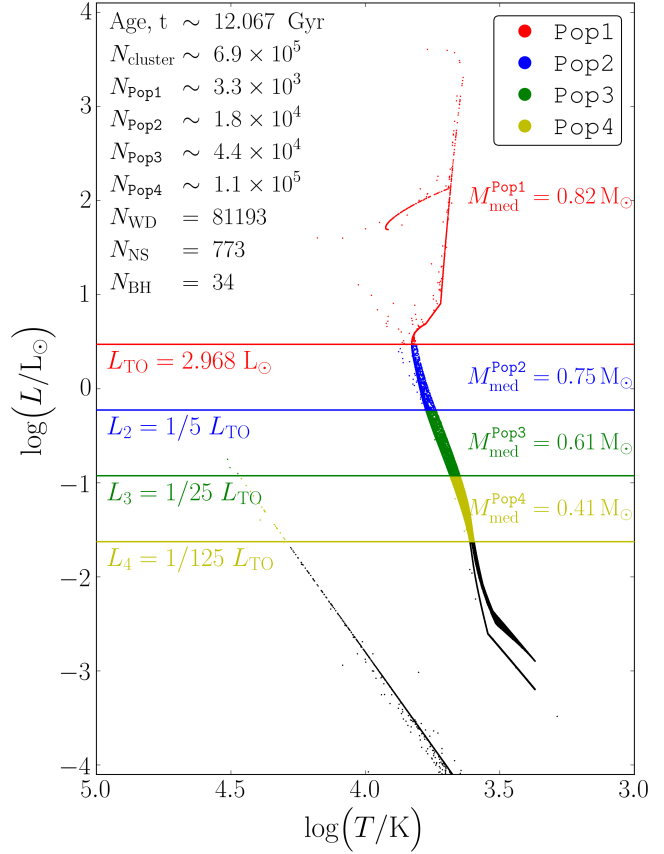}
\caption{Example Hertzsprung-Russell diagram showing the four stellar populations used to measure mass segregation in a typical \texttt{CMC} model at $12\,\gyr$ ($N=8\times 10^5$, $r_v= 1\,\pc$, $R_{\rm{gc}}=8\,\kpc$, and $\metal=0.01\,\metal_\odot$). Each data point represents a single or binary star (all binaries are considered unresolved). The highest-mass population (\popone, in red) encompasses all stars above the MS turnoff (delineated in red), which is defined as the luminosity corresponding to the highest temperature on the MS (excluding blue stragglers). The three lower-mass populations (\poptwo\ in blue, \popthree\ in green, and \popfour\ in yellow) are evenly spaced in log scale along the MS with lower boundaries delineated in the corresponding colors. The median masses for all populations are shown in the figure. Relevant cluster properties at the time of this snapshot such as $N$, $\nbh$, and the respective numbers in each stellar population are also included. Defining stellar populations this way ensures a high number of stars in each population, with highest sample sizes for dimmer populations.}
\label{f:1}
\end{figure}

In general, quantifying $\Delta$ in a star cluster requires comparing the radial distributions of multiple stellar populations sufficiently different in their average masses \citep[e.g.,][]{Goldsbury2013}. While stellar mass is not directly measured in real clusters, stellar luminosity is, and can be used as a proxy for mass, especially for main sequence (MS) stars \citep[e.g.,][]{Hansen1994}. As in W1, we anchor our population definitions to the location of the MS turnoff (MSTO), the most prominent feature on a color-magnitude or Hertzsprung-Russel diagram (\autoref{f:1}). Defining the MSTO at $L=\Lto$, where the MS stars (excluding blue stragglers) exhibit the highest temperature, population bounds are then established as fractions of $\Lto$. While these details are unchanged from W1, we have upgraded the specific population choices used to measure $\Delta$.

In W1, we sought to maximize the signal strength in $\Delta$ by choosing two populations with characteristic masses (luminosities) as different as possible while still ensuring that the lighter population is bright enough to be easily observable in the MWGCs. In addition, both populations must contain large enough numbers of stars to limit statistical scatter. Under these constraints, we chose a high-mass population containing all stars with $L>\Lto$ and a low-mass population consisting of MS stars with $\Lto/125\leq L\leq \Lto/25$. While these population choices (\popone\ and \popfour\ in \autoref{f:1}) maximized the magnitude of $\Delta$ while ensuring relatively large observable population sizes, reducing statistical scatter compared to other choices in previous studies \citep[e.g., blue stragglers;][]{Peuten2016,Alessandrini2016}, they were not free from drawbacks \citep{deVita2019}. Specifically, \popone\ contains far fewer stars than any of the three MS populations, introducing higher statistical scatter than strictly necessary. Furthermore,
an extreme luminosity difference between populations can cause them to suffer from large discrepancies in observational incompleteness, in which dim stars are washed out by bright neighbors. As shown in W1, difference in the radially-dependent incompleteness between populations can introduce significant uncertainty in the measured $\Delta$, and by extension, the inferred number of BHs.

For example, the median masses for \popone\ and \poptwo\ are $0.82\,\msun$ and $0.75\,\msun$, respectively, a minor gap (\autoref{f:1}). Meanwhile, the stellar luminosity in \popone\ spans nearly 3 orders of magnitude. Independent of our choice for the other population, \popone's inclusion in the $\Delta$ calculation balloons the incompleteness difference between populations, resulting in increased uncertainty. In contrast, the three MS populations (\poptwo, \popthree, \popfour) differ much more significantly in their median mass with comparatively small variation in their typical luminosities. In this work, we therefore use only these MS populations to compute $\Delta$ and ignore \popone.

\subsection{Quantifying Mass Segregation} \label{S:delta_def}
Having chosen three distinct MS populations, we compute the mass segregation, $\Delta$, between any pair of them using both parameters introduced in W1. The first, $\Deltarfifty^{ij}$, is the difference in median cluster-centric distance between $\pop i$ and $\pop j$. The second, $\Deltaa^{ij}$ is the difference in area under the two populations' cumulative radial distributions. In both cases, the cluster-centric radial distances used are sky-projected and normalized by the cluster's sky-projected half-light radius to make $\Delta$ unitless. Mathematical expressions and graphical representations of these mass segregation parameters are given in Section 2.3 of W1.

\subsection{$\Delta$ vs $\nbh/\ncluster$ ($\mbh/\mcluster$) in models: Effects of cluster properties} 
\label{S:model_trend}

As introduced in W1, there exists a strong anti-correlation between the ratio of BHs to total stars retained in a cluster ($\nbh$/$\ncluster$) and the cluster's measured mass segregation ($\Deltarfifty^{ij}$ and $\Deltaa^{ij}$). The anti-correlation is due to BH burning, in which BHs mass-segregate to the core and provide energy to passing stars via two body relaxation, pushing those stars farther out into the cluster. On average, the most massive stars gain proportionally more energy since they are distributed closer to the BH-core than less massive stars. Hence, clusters with more numerous (more massive) BH populations in their core display reduced mass segregation.

\begin{figure*}
\epsscale{1.175}
\plotone{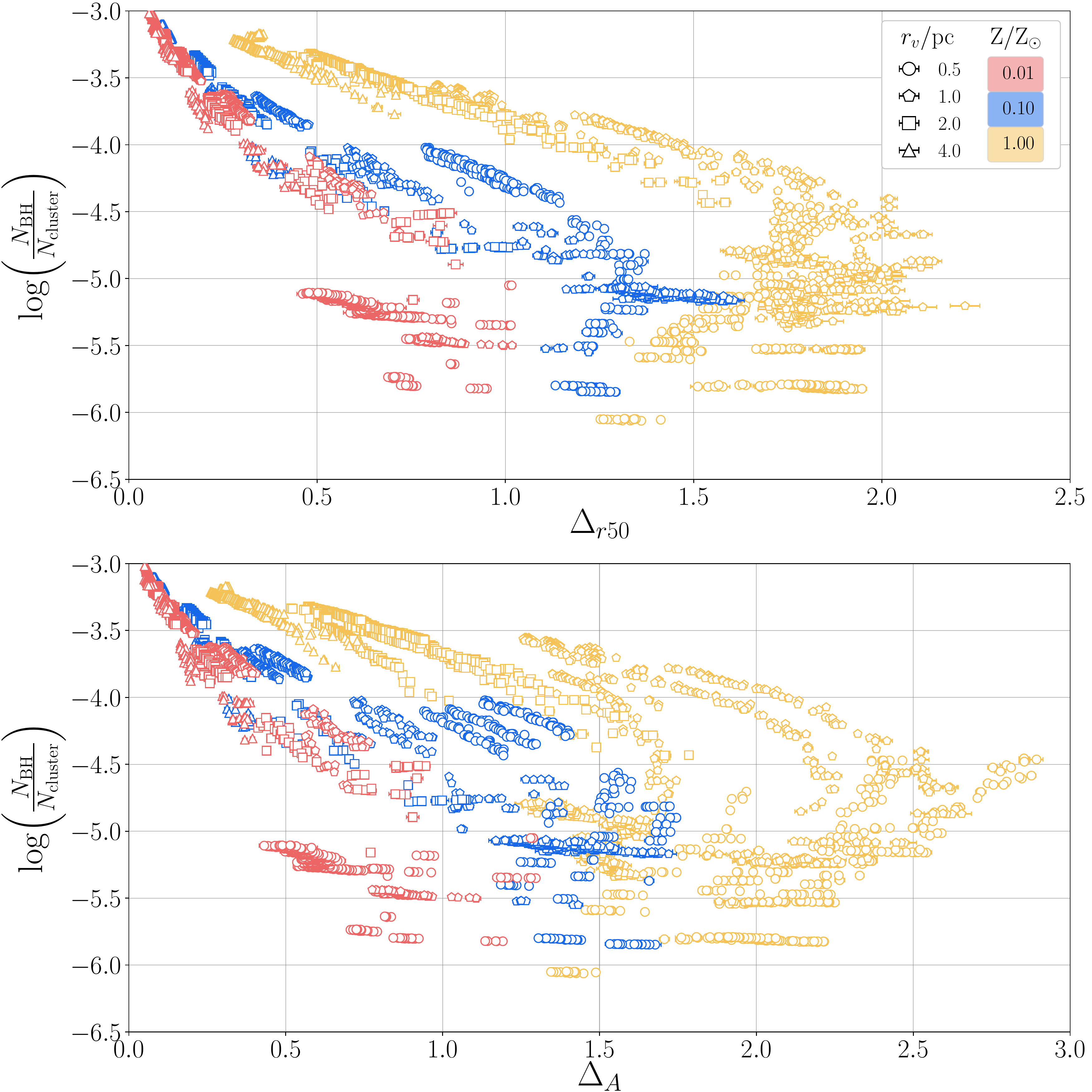}
\caption{Number of retained BHs per cluster star $\nbh/\ncluster$ vs mass segregation parameters $\Deltarfifty$ (top) and $\Deltaa$ (bottom), calculated between \poptwo\ and \popfour\ under a standard radial limit of $\rlim=\rhl$ for all model snapshots with $9\leq t \leq 13\,\gyr$. Each data point represents the mean $\Delta$ across $10$ realizations of 2D projections of all stars' radial positions in a snapshot. Uncertainty bars represent the standard deviations within these realizations. Color distinguishes models by stellar metallicity $\metal/\metal_{\odot}$, shape by virial radius $r_v$ (see legend). A clear anti-correlation between $\nbh/\ncluster$ and $\Delta$ is apparent, especially when models of particular $\metal$ are considered separately. In so doing, it is evident that higher $\metal$ results in higher $\Delta$ for any given $\nbh/\ncluster$. This occurs because BH masses decrease as $\metal$ increases. Thus, to effect the same level of quenching of $\Delta$, a higher $\nbh/\ncluster$ is needed. The slight backwards curvature in the trend peaking around $\nbh/\ncluster \sim 10^{-5}$ is due to the nondimensionalization of $\Delta$, where the half-light radius normalization factor increases faster than $\Delta$ at higher dynamical ages (generally, following the curved trajectories down from upper left). Finally, keep in mind that the number of snapshots per simulation varies significantly based on relaxation time, so this figure -- unlike our calculations -- does $not$ weight simulations equally.}
\label{f:2}
\end{figure*}

In \autoref{f:2}, we show the $\Delta$-$\nbh/\ncluster$ anti-correlation across all $7,355$ model snapshots with ages between $9$-$13\,\gyr$, colored by metallicity and using a standard radial limit of $\rlim = \rhl$. (i.e. Only stars within the model clusters' half-light radii are used when measuring $\Delta$ for this figure, a constraint motivated by field-of-view limits when observing real clusters; see \autoref{S:observational_results}). In the top panel, $\Deltarfifty^{24}$ ($\Deltarfifty$ between \poptwo\ and \popfour) is used for $\Delta$, while the lower panel uses $\Deltaa^{24}$. Uncertainty bars represent the standard deviation across the 10 randomized 2D projections (`views') of each cluster snapshot.

Though not shown, plots of $\mbh/\mcluster$ versus $\Delta$ are practically indistinguishable from \autoref{f:2}, except with a y-axis range of $\log(\mbh/\mcluster) \in [-1.3,-5.3]$. Other pairings of the four populations in \autoref{f:1} to measure $\Delta$ also result in very similar anti-correlations, though wider spread is indeed apparent whenever \popone\ is used, for the reasons discussed earlier.

With both more models and much fuller coverage of the space of initial cluster parameters characterizing observed MWGCs ($N$, $r_v$, $\metal$, and $R_{\rm{gc}}$), the anti-correlation extends to larger mass segregation and to an order-of-magnitude lower $\nbh/\ncluster$ than in W1. The metallicity dependence of the trend is also more explicit. The higher the $\metal$, the lower the mass of the BHs produced, so higher-$\metal$ clusters need higher $\nbh/\ncluster$ to quench $\Delta$ to the same degree as a lower-$\metal$ cluster.

Other parameters contribute less visibly to the spread in the trend, primarily through their impact on dynamical age, which increases from upper left to lower right along the trend. Specifically, a detailed model-to-model examination reveals that virial radius ($r_v$) has the largest impact on a snapshot's location along the trend at any given physical time. Clusters with lower initial $r_v$ relax faster, making them dynamically older at late times than GCs with higher $r_v$. Since $\Delta$ correlates with and $\nbh$ anti-correlates with dynamical age, the models with lowest $r_v$ appear at the bottom right of each panel in \autoref{f:2}. Initial $N$ also affects the relaxation timescale of a cluster. Thus, the least massive clusters are also dynamically the oldest at the same physical time. These low-mass clusters tend to be at the bottom right. Similarly, all else being fixed, as a cluster gets older, it moves down and right along the trend, albeit to a lesser degree than movement from $N$ or $r_v$ variation, since the age range used here is narrow ($9$-$13\,\gyr$) compared to lifetimes of typical GCs. For an average model, $\nbh/\ncluster$ drops by 0.5 dex between the $9$ and $13\,\gyr$ snapshots. Finally, increasing Galactocentric distance ($R_{\rm{gc}}$) slightly increases $\Delta$ but has little impact on $\nbh/\ncluster$, shifting snapshots left-to-right in the figure. This occurs because clusters farther from the Galactic center experience lower tidal forces, increasing the cluster's tidal radius (boundary) and making it harder for stars to escape the cluster. As will be discussed in the next section, limiting the radial extent of the stellar populations used to measure $\Delta$ decreases $\Delta$.

%
\subsection{Measuring Mass Segregation in Observed Clusters} \label{S:observational_results}
%
%

\begin{figure}
\epsscale{1.175}
\plotone{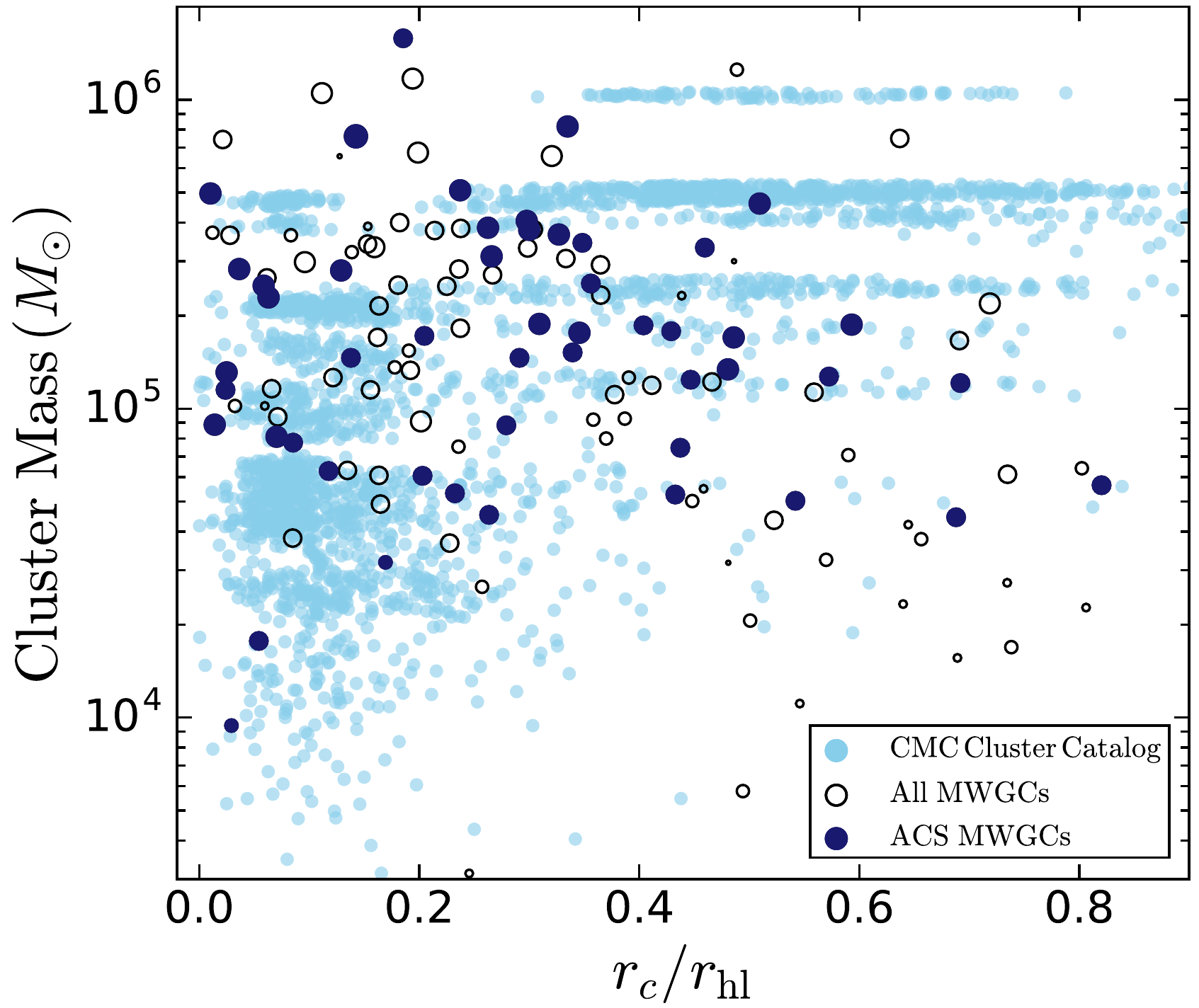}
\caption{Total cluster mass versus $r_c/r_{\rm{hl}}$ for all models in the \texttt{CMC} Cluster Catalog \citep[light blue; from][]{Kremer2020}, the 50 ACS Survey clusters studied in this analysis (dark blue), and the complete set of MWGCs \citep[open circles; from][]{Baumgardt2018}. The size of each dark blue and open circle corresponds to the integrated V-band magnitude of the corresponding GC \citep[][2010 edition]{Harris1996}.
}
\label{f:3}
\end{figure}

To measure $\Delta$ in real clusters, we use the ACS Survey of MWGCs \citep{Sarajedini2007}. Compiled using the wide-field channel of the \textit{Hubble Space Telescope's} Advanced Camera for Surveys (ACS), this resource catalogs stars within the central $4\arcmin\times4\arcmin$ of 71 MWGCs and exists as an online database of stellar coordinates and calibrated photometry in the ACS VEGA-mag system \citep{Sirianni2005}. Construction of the database, which may be accessed publicly at \url{http://www.astro.ufl.edu/~ata/public\_hstgc}, is fully detailed by \cite{Anderson2008}.

Using the observed stellar data, we construct the same four turnoff-anchored populations as described above for the models. The exact procedure used for constructing observed populations is fully described and illustrated in Section 4 of W1. A couple important steps are worth highlighting. First, since the ACS field-of-view (FOV) is a rhombus covering only the central-most region of each GC, using raw ACS stellar data to construct the populations will introduce a radial bias in observed $\Delta$ when comparing to $\Delta$ in the models, which have effectively unlimited FOVs. We therefore establish a radial limit ($\rlim$) for each ACS-observed MWGC as the radius of the largest circle inscribable in its FOV. We then measure $\Deltarfifty$ and $\Deltaa$ between each pair of the observed cluster's three MS populations, including only stars within that specific MWGC's radial limit. When later applying the $\Delta-\nbh,\mbh$ correlations from the models to predict $\nbh$ and $\mbh$ in that MWGC, we utilize model data that has been radially-limited to match the observed $\rlim$. For our set of 50 MWGCs, $\rlim \in [0.52,3.48]\times \rhl$.

Second, we found in W1 that observational incompleteness significantly impacts observed $\Delta$ measurements. Correcting for incompleteness is even more critical in this broader survey of MWGCs, as there are more extreme examples with low completeness, especially for core-collapsed clusters in which dim stars, even relatively far from the cluster center, are almost entirely washed out by the bright ones. Even in non-core-collapsed clusters, changes in $\Delta$ of order 50\% are common after correcting for incompleteness.

We correct for observational incompleteness in each ACS-observed cluster using the procedure described in Section 4.3 of W1. The procedure relies on artificial star files included in the ACS Globular Cluster Survey, discussed in Section 6 of \cite{Anderson2008}. In short, \citet{Anderson2008} inject $10^5$ artificial point spread functions (stars) into each of their raw ACS images, using the fraction of recovered to total injected stars as a proxy for true observational completeness. Using this data, we compute a kernel density estimate (KDE) of this `completeness fraction' as a function of cluster-centric distance $r$ and apparent V-band magnitude $m_V$. Using KDEs reduces much of the uncertainty, coarseness, and bias by leveraging the full statistical power of all $10^5$ artificial stars compared to other methods, such as $r,m_V$-binning. Each observed (non-artificial) star is assigned a completeness fraction based on its location in the $(r,m_V)$ space. While calculating $\Delta$, we randomly under-sample the stars that are more complete compared to those that are less complete by a factor given by the ratio of their completeness fractions. We repeat this exercise $10^3$ times to find a distribution of the measured values of $\Delta$. Thus, because of completeness differences between stars of different $r$ and $m_V$ in a MWGC, instead of a single value of $\Delta$, we obtain a distribution representative of the observational uncertainties for that cluster. This process and its importance are discussed in more detail in Section 4.3 of W1. We find that the uncertainties on $\Delta$ take the form of Gaussian probability density functions (PDFs) with typical $1\sigma$ uncertainty of order $10\%$ or less among the 50 ACS clusters we analyse.

In W1, we limited our analysis to 47\,Tuc, M\,10, and M\,22 -- all known to contain candidate stellar-mass BHs \citep[e.g.,][]{Strader2012,Shishkovsky2018,Miller-Jones2015,Bahramian2017}. In this full survey, we predict $\nbh$ and $\mbh$ for 50 of the ACS Survey's 71 MWGCs. We do not analyze 21 GCs from the ACS catalog for varied reasons. Eight (IC04499, PAL2, PAL15, PYXIS00, RUPR106, and NGCs 0362, 6426, 7006) are excluded because the catalog does not include the necessary information (artificial star files) for performing incompleteness corrections. Three MWGCs (NGCs 6362, 6388, 6441) are excluded because their artificial star data are incomplete, two ($\omega$ Cen and NGC 6121) are excluded because their FOVs do not extend to at least 0.5 $\rhl$, and one (NGC 6496) is excluded because its FOV is half-size and triangular rather than rhomboidal. The remaining seven of the survey's non-NGC clusters (ARP2, E3, LYNGA7, PAL1, PAL2, TERZAN7, and TERZAN8) are excluded because of their general status as outliers relative to the bulk of the MWGCs and the limited coverage by our models of their (lower-right) region of the mass vs. $r_c/\rhl$ parameter space, seen in \autoref{f:3}. In this figure, we compare the cluster properties of the selected 50 ACS Survey clusters to the full population of MWGCs \citep[taken from][]{Baumgardt2018} as well as the models from the \texttt{CMC} Cluster Catalog. The figure shows that both the \texttt{CMC} models and the selected ACS clusters cover a very similar parameter space, providing confidence in our analysis. In addition, the analyzed clusters span roughly the entire parameter space for all MWGCs, indicating that results from this study are likely representative of the entire population of MWGCs.

%
\subsection{Comparing Models to Observations} \label{S:mseg_comp}
%

\begin{figure}
\epsscale{1.175}
\plotone{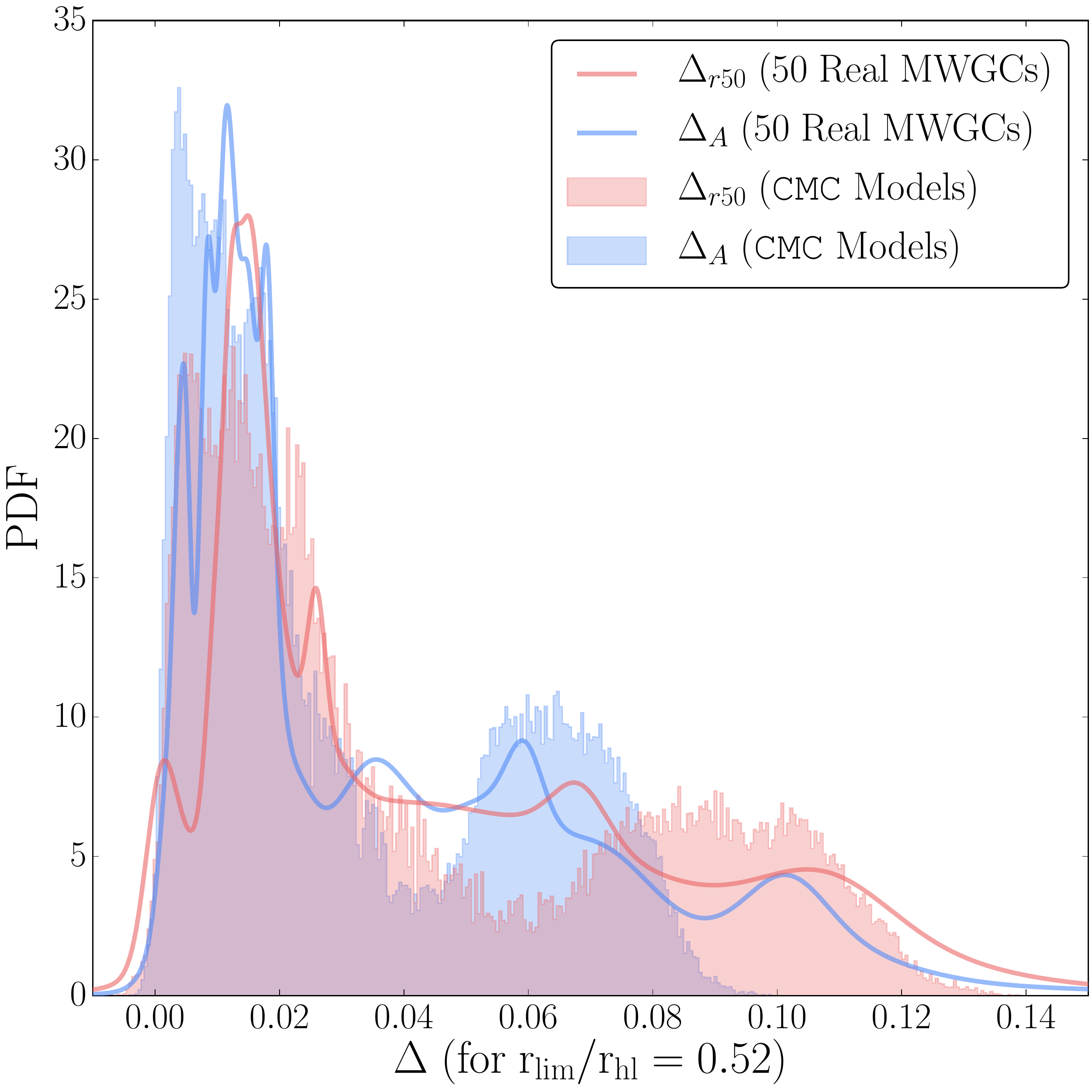}
\caption{Normalized mass segregation distributions across all model snapshots (weighted equally; filled histograms) versus the distributions across the 50 observed MWGCs from the ACS Survey (solid curves). The ACS curves are simply normalized sums of the Gaussian PDFs representing the uncertainties on each $\Delta$ measurement after correcting for incompleteness. Here, $\Deltarfifty$ (red) and $\Deltaa$ (blue) are measured between \poptwo\ and \popfour. To compare all modeled/observed GCs evenly, only stars within 0.52 projected half-light radii of each cluster center were included in the $\Delta$ computations, corresponding to the narrowest radial limit among the 50 ACS GCs analyzed. The close match between the \texttt{CMC} and ACS $\Delta$ distributions demonstrates that our models accurately capture the state of mass segregation in MWGCs, even while not having been specifically tuned to do so.}
\label{f:4}
\end{figure}

First, we compare the $\Delta$ measured in our models with those measured in the 50 MWGCs we analyze (\autoref{f:4}). Here, the $\Delta$ distribution from our models is shown as a simple normalized histogram whereas the observed distribution is obtained by summing the 50 MWGCs' individual incompleteness-corrected $\Delta$ distributions measured in \autoref{S:observational_results}. In both cases we use the same radial limit ($\rlim=0.52\,\rhl$) to ensure a fair comparison. The excellent similarity between the observed and model $\Delta$ distributions bolsters our belief that the parameter space covered in our models is representative of the MWGCs and using the $\Delta$--$\nbh$ correlation in our models to estimate $\nbh$ for MWGCs is appropriate. For further reference, the modes and $1\sigma$ uncertainties on $\Delta$ for all MWGCs analyzed are listed in \autoref{T:raw_results_dr} ($\Deltarfifty^{24}$) and \autoref{T:raw_results_dA} ($\Deltaa^{24}$).

The $\Delta$ distributions from our models and the MWGCs are remarkably similar, not only in range but also in rough shape. This is especially noteworthy considering how strongly the magnitude of $\Delta$ depends on the imposed radial limit and the incompleteness correction. For example, the tight limit, $\rlim = 0.52\,\rhl$, in \autoref{f:4} reduces the typical, unlimited value of $\Delta$ by a full order of magnitude from $\Delta \sim 0.1$ to $\Delta \sim 0.01$. Such a comparatively close match between the model and observed $\Delta$ distributions therefore provides strong evidence that our \texttt{CMC} models at $9$--$13\,\gyr$ accurately capture the state of mass segregation in MWGCs. Furthermore, this similarity is achieved without having specifically tuned the models to match observed mass segregation; instead, the match derives simply from the observationally-motivated grid of chosen initial conditions. The match also demonstrates that our main sequence population-based method of measuring mass segregation is both highly robust and adaptable to significant FOV limitations.

Finally, it is worth pointing out that while the $\Delta$ distribution from the models appears strongly bimodal, even tetramodal, this is merely an artifact of the model grid. The four different initial virial radii divide the model set into four subsets with different initial relaxation times and accordingly divergent levels of mass segregation at late times. Variations in initial $N$ and snapshot age smooth out the resulting spectrum in $\Delta$, but the discreteness of the model grid should \textit{not} be mistaken for a fundamental physical phenomenon. In turn, however, the ACS-observed $\Delta$ distribution exhibits a similarly strong peak to the model distribution at $\Delta \sim 0.02$. This specific value is unimportant, as it depends on the radial limit, but the peak's existence does appear to be a statistically significant feature of the true mass segregation distribution for MWGCs. This peak is representative of dynamically young clusters that have yet to undergo core-collapse and retain many BHs. The tail in the distribution is also likely a true feature, representative of dynamically older clusters that have depleted most of their BHs. Together, these features likely reflect a common distribution of initial cluster size and mass in the Milky Way, contaminated by numerous dynamically-younger GCs accreted from nearby dwarf galaxies \citep[e.g.,][]{Searle1978,Mackey2004,Kruijssen2019}.

%
\section{Predicting the Number and Mass of Retained Black Holes in Observed GCs} \label{S:nbh_predictions}
%

\begin{figure*}[htb!]
\epsscale{1.177}
\plotone{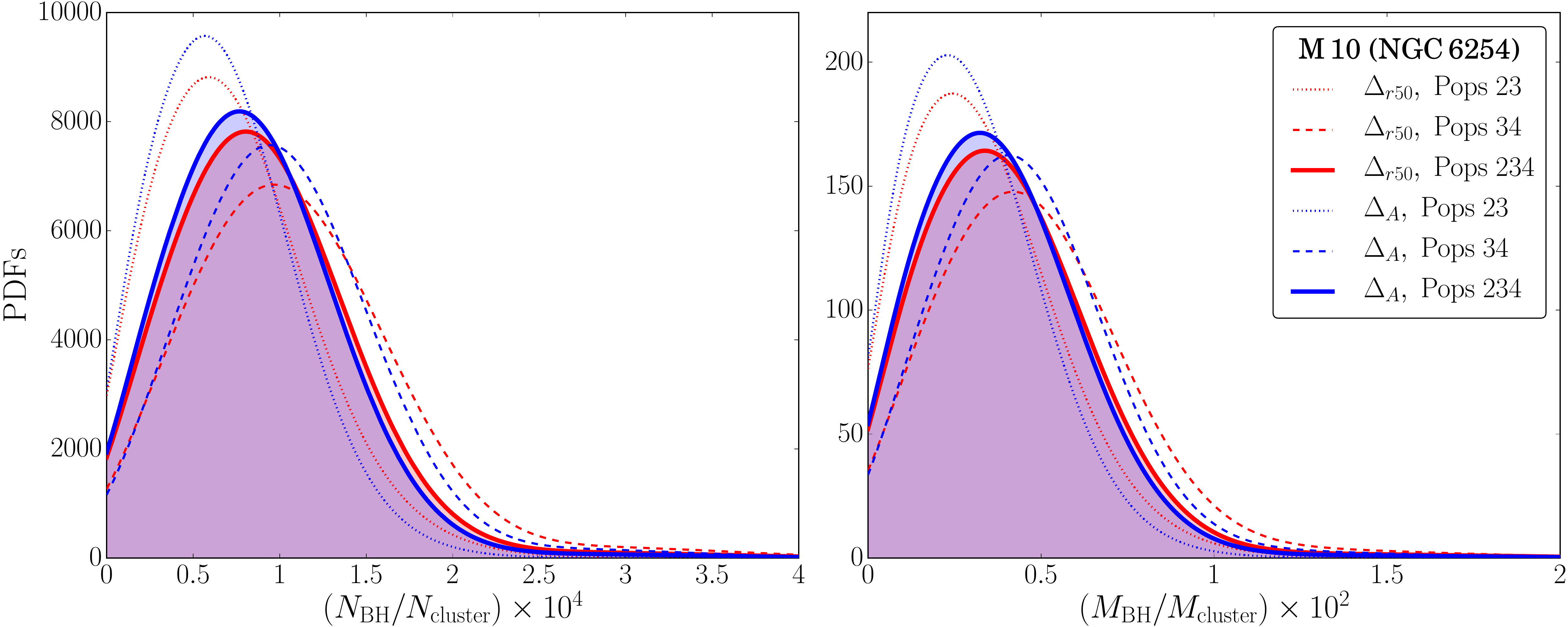}
\caption{Example probability density functions (PDFs) for $\nbh/\ncluster$ and $\mbh/\mcluster$ in the globular cluster M\,10. The complete figure set (50 images) for all MWGCs studied in this paper is available in the online journal. In all cases, the final PDFs are computed by plugging the measured, completeness-corrected mass segregation estimate ($\Delta$ as a Gaussian PDF) into a multivariate kernel density estimate (KDE) of the $\Delta$-$\nbh/\ncluster$ ($\Delta$-$\mbh/\mcluster$) parameter space from the models (e.g., \autoref{f:2}, except in linear scale and often with a different radial limit). Predictions made using $\Deltarfifty$ are colored in red while those made using $\Deltaa$ are colored in blue. The dotted distributions correspond to predictions based only on the observed mass segregation between \poptwo\ and \popthree\ ($\Delta^{23}$), the dashed distributions only on $\Delta^{34}$ between \popthree\ and \popfour. The solid, filled distributions are based on both $\Delta^{23}$ \textit{and} $\Delta^{34}$ by adding an extra dimension to the KDE (see text). The modes, $1\sigma$, and $2\sigma$ confidence intervals of these solid distributions are our final predictions, presented in \autoref{T:raw_results_dr}. Finally, note that the x-axis tick marks are less than one, and have merely been re-scaled by the indicated factors for cleaner labeling.}
\label{f:5}
\end{figure*}

\noprint{\figsetstart}
\noprint{\figsetnum{5}}
\noprint{\figsettitle{BH Distributions for 50 MWGCs}}

\figsetend

We now derive PDFs for $\nbh/\ncluster$ and $\mbh/\mcluster$ retained in all 50 of the MWGCs analyzed, as inferred from the appropriate radially-limited, metallicity-matched model sets and measured $\Deltarfifty$ or $\Deltaa$ (hereon referred to jointly as just $\Delta$). Unlike in W1, however, we have multiple different measurements of $\Delta$ for each cluster -- one for each pairing of the three MS populations (i.e. $\Delta^{23}$, $\Delta^{24}$, $\Delta^{34}$). In order to combine the measurements into a single prediction, we compute Gaussian kernel density estimates (KDEs) of the $\Delta^{23}-\Delta^{34}-\nbh/\ncluster$ ($\Delta^{23}-\Delta^{34}-\mbh/\mcluster$) space. For those less familiar with such a method, multivariate KDEs are mere generalizations of the standard univariate KDE, a non-parametric way to estimate the PDF of a random variable. We use the standard \texttt{gaussian\_kde} function included in \texttt{SciPy}'s statistics package.

For each MWGC, we select only models with $\metal$ closest to its observed metallicity \citep[][2010 edition]{Harris1996} determined on the basis of simple logarithmic binning. Specifically, BH predictions for MWGCs with $\metal/\metal_\odot < 0.033$ are based only on the models with $\metal/\metal_\odot=0.01$, predictions for MWGCs with $0.033 < \metal/\metal_\odot < 0.067$ are based on models with $\metal/\metal_\odot=0.01$ and $0.1$, and those for MWGCs with $0.067 < \metal/\metal_\odot < 0.133$ incorporate models with $\metal/\metal_\odot=0.1$. For only the few MWGCs with $\metal/\metal_\odot>0.133$ do we use the models with $\metal/\metal_\odot= 0.1$ and 1. In all cases, we exclude $\Delta^{24}$ as a fourth axis in the KDE because it is simply the sum of $\Delta^{23}$ and $\Delta^{34}$ and hence is not an independent additional axis.  These trivariate distributions are then used to infer the expected number (total mass) of retained BHs in each GC using the following procedure.

For each GC, we evaluate the above 3D PDF (from the \textit{models}) on a grid of points spanning the $3\sigma$ confidence intervals (CIs) of the observed $\Delta^{23}$ and $\Delta^{34}$, and from $\nbh/\ncluster$ ($\mbh/\mcluster$) = 0 to twice the maximum $\nbh/\ncluster$ ($\mbh/\mcluster$) seen in our models. The sample points are spaced evenly in linear scale along all three axes -- $\Delta^{23}$, $\Delta^{34}$, and $\nbh/\ncluster$ ($\mbh/\mcluster$) -- with a respective grid size of $15\times15\times1001$ sample points. This resolution is high enough to ensure that an order of magnitude resolution increase along each axis changes the final mode, $1\sigma$ and $2\sigma$ CIs on $\nbh/\ncluster$ ($\mbh/\mcluster$) in the third significant digit, at the most. In this grid form, the 3D PDF from the models is then convolved with both 1D PDFs characterizing the uncertainty on $\Delta^{23}$ and $\Delta^{34}$ observed in the MWGC (see \autoref{S:observational_results}). The resulting convolution is integrated numerically via Simpson's rule (also implemented in \texttt{SciPy}) along the $\Delta^{23}$ and $\Delta^{34}$ axes. The integral is then normalized to obtain the final 1D PDFs for $\nbh/\ncluster$ ($\mbh/\mcluster$). These distributions (filled, solid curves) are exemplified for the case of NGC 6254 (M\,10) in \autoref{f:5}, based on both $\Deltarfifty$ (red) and $\Deltaa$ (blue). Note that these final PDFs are equivalent to those shown in Figure 10 of W1, just with a different KDE formulation from the one used in that paper (namely, the addition of an extra axis to the KDE and convolution of the raw KDE with the observed $\Delta$ PDFs rather than Monte Carlo sampling to reduce computational cost). Versions of \autoref{f:5} for all 50 MWGCs analyzed in this study are available in the online journal. The corresponding modes, $1\sigma$, and $2\sigma$ confidence intervals on $\nbh/\ncluster$ ($\mbh/\mcluster$) for each GC are reported in \autoref{T:raw_results_dr} for predictions based on $\Deltarfifty$. For the nearly identical results based on $\Deltaa$, see \autoref{T:raw_results_dA} in the Appendix. Because $\Deltarfifty$ is simpler to calculate, we recommend using it in the future rather than $\Deltaa$, but we recognize that some observers seem to prefer $\Deltaa$ \citep[e.g.,][]{Alessandrini2016}.

\floattable
\begin{deluxetable*}{lccccc|rrrrr|rrrrr}
\centerwidetable
\tabletypesize{\scriptsize}
\tablecolumns{16}
\tablewidth{0pt}
\tablecaption{Cluster Properties and Raw Computational Results Based on $\Deltarfifty$}
\tablehead{ & & & & & & & & & & & & & & & \vspace{-4pt}\\
   \multirow{2}{*}{Cluster} & \multirow{2}{*}{$\displaystyle{\frac{\rlim}{\rhl}}$} & \multirow{2}{*}{$\displaystyle{\frac{\mcluster}{\lcluster}}$} & \multicolumn{2}{c}{$\mcluster/(10^3\cdot\msun)$} & $\Deltarfifty^{24}$ & \multicolumn{5}{c}{$(\nbh/\ncluster)\cdot 10^5$} & \multicolumn{5}{c}{$(\mbh/\mcluster)\cdot 10^5$} \vspace{0.00cm} \\
   & & & Baumgardt & Harris & $\pmsigma$ & $-2\sigma$ & $-1\sigma$ & Mode & $+1\sigma$ & $+2\sigma$ &
   $-2\sigma$ & $-1\sigma$ & Mode & $+1\sigma$ & $+2\sigma$}
\startdata
NGC 0104 (47Tuc)  & 0.55 & 1.77 &  779 & 1000 & $0.062 \pm 0.009$ &     0 &  0.41 & \textbf{2.75} &  6.87 &  12.1 &    0 &   13 & \textbf{117} &  302 & 555  \\
NGC 0288          & 0.77 & 2.39 &  116 &   87 & $0.015 \pm 0.002$ &  1.15 &  6.17 & \textbf{11.3} &  16.4 &  22.5 &   39 &  263 & \textbf{512} &  766 & 1048 \\
NGC 1261          & 2.45 & 2.12 &  167 &  225 & $0.035 \pm 0.014$ &  1.05 &   6.0 & \textbf{11.7} &  18.1 &  24.3 &   27 &  249 & \textbf{506} &  771 & 1043 \\
NGC 1851          & 3.48 & 2.02 &  302 &  367 & $0.104 \pm 0.030$ &     0 &   0.5 & \textbf{3.22} &  7.75 &  14.0 &    0 &   19 & \textbf{139} &  344 & 624  \\
NGC 2298          & 1.70 & 0.46 &   12 &   57 & $0.032 \pm 0.007$ &     0 &  1.09 & \textbf{4.43} &  8.84 &  14.2 &    0 &   34 & \textbf{177} &  401 & 690  \\
NGC 2808          & 2.25 & 1.64 &  742 &  975 & $0.058 \pm 0.013$ &     0 &  1.84 & \textbf{5.87} &  10.3 &  15.1 &    0 &   56 & \textbf{212} &  408 & 631  \\
NGC 3201          & 0.57 &  2.4 &  149 &  163 & $0.013 \pm 0.003$ &     0 &  2.24 & \textbf{13.7} &  27.2 &  62.8 &    0 &   67 & \textbf{602} & 1246 & 3265 \\
NGC 4147          & 3.48 & 1.51 &   33 &   50 & $0.042 \pm 0.015$ &     0 &  0.65 & \textbf{3.55} &  8.44 &  14.3 &    0 &   28 & \textbf{166} &  411 & 713  \\
NGC 4590 (M68)    & 1.15 & 2.02 &  123 &  152 & $0.014 \pm 0.003$ &     0 &  2.57 & \textbf{6.81} &  11.1 &  15.4 &    0 &   84 & \textbf{275} &  488 & 722  \\
NGC 4833          & 0.73 & 0.84 &  247 &  317 & $0.014 \pm 0.002$ &     0 &  4.51 & \textbf{12.6} &  21.3 &  42.3 &    0 &  164 & \textbf{547} &  962 & 2090 \\
NGC 5024 (M53)    & 1.29 & 1.59 &  380 &  521 & $0.043 \pm 0.010$ &     0 &  2.19 & \textbf{6.69} &  11.5 &  16.8 &    0 &   78 & \textbf{277} &  515 & 787  \\
NGC 5053          & 0.68 & 1.66 &   57 &   87 & $0.012 \pm 0.001$ &  12.1 &  17.3 & \textbf{48.7} &  61.1 &  91.9 &  521 &  759 & \textbf{2443} & 3052 & 4588 \\
NGC 5272 (M3)     & 0.77 & 1.56 &  394 &  610 & $0.047 \pm 0.009$ &     0 &  0.48 & \textbf{3.22} &  8.07 &  14.2 &    0 &   18 & \textbf{149} &  390 & 707  \\
NGC 5286          & 2.25 & 1.41 &  401 &  536 & $0.099 \pm 0.019$ &     0 &  0.23 & \textbf{2.53} &  6.75 &  12.5 &    0 &   10 & \textbf{123} &  332 & 622  \\
NGC 5466          & 0.77 & 1.13 &   46 &  106 & $0.001 \pm 0.002$ &   5.1 &  11.4 & \textbf{20.3} &  42.3 &  73.5 &  153 &  431 & \textbf{928} & 2001 & 3816 \\
NGC 5904 (M5)     & 1.00 & 1.52 &  372 &  572 & $0.040 \pm 0.008$ &     0 &   1.9 & \textbf{6.1} &  11.0 &  16.7 &    0 &   67 & \textbf{243} &  463 & 724  \\
NGC 5927          & 1.62 & 2.61 &  354 &  228 & $0.015 \pm 0.011$ &  3.43 &  9.74 & \textbf{17.4} &  24.4 &  38.5 &  106 &  373 & \textbf{706} & 1038 & 1726 \\
NGC 5986          & 1.70 & 2.45 &  301 &  406 & $0.021 \pm 0.018$ &  0.35 &  7.21 & \textbf{14.1} &  21.2 &  33.0 &    0 &  285 & \textbf{613} &  955 & 1523 \\
NGC 6093 (M80)    & 2.89 & 1.43 &  249 &  335 & $0.120 \pm 0.016$ &     0 &  0.44 & \textbf{3.03} &  7.73 &  13.7 &    0 &   18 & \textbf{143} &  373 & 672  \\
NGC 6101          & 1.62 &  3.0 &  127 &  102 & $0.002 \pm 0.003$ &  29.4 &  40.9 & \textbf{49.1} &  74.7 &  93.0 & 1376 & 1966 & \textbf{2402} & 3918 & 4630 \\
NGC 6144          & 1.00 & 0.54 &   45 &   94 & $0.018 \pm 0.003$ &  1.27 &  7.82 & \textbf{14.7} &  21.8 &  39.2 &    0 &  317 & \textbf{659} & 1008 & 1888 \\
NGC 6171 (M107)   & 1.00 & 2.16 &   87 &  121 & $0.018 \pm 0.004$ &  0.03 &  5.23 & \textbf{13.2} &  24.8 &  44.2 &    0 &  207 & \textbf{589} & 1096 & 2078 \\
NGC 6205 (M13)    & 1.00 & 2.61 &  453 &  450 & $0.021 \pm 0.006$ &     0 &  6.73 & \textbf{14.1} &  21.6 &  38.1 &    0 &  260 & \textbf{615} &  984 & 1864 \\
NGC 6218 (M12)    & 0.99 & 1.27 &   87 &  144 & $0.016 \pm 0.003$ &     0 &  6.13 & \textbf{12.6} &  20.5 &  37.3 &    0 &  269 & \textbf{588} &  950 & 1742 \\
NGC 6254 (M10)    & 0.89 & 1.94 &  184 &  168 & $0.022 \pm 0.003$ &     0 &  3.14 & \textbf{8.05} &  13.2 &  18.8 &    0 &  112 & \textbf{338} &  584 & 876  \\
NGC 6304          & 1.29 & 1.37 &  277 &  142 & $0.061 \pm 0.025$ &     0 &  2.78 & \textbf{12.6} &  18.5 &  27.3 &    0 &   82 & \textbf{262} &  455 & 655  \\
NGC 6341 (M92)    & 1.70 & 1.81 &  268 &  329 & $0.077 \pm 0.023$ &     0 &  0.65 & \textbf{3.47} &  8.28 &  14.0 &    0 &   26 & \textbf{161} &  402 & 700  \\
NGC 6352          & 0.85 & 2.47 &   94 &   66 & $0.028 \pm 0.004$ &     0 &  2.78 & \textbf{7.53} &  12.5 &  20.6 &    0 &  104 & \textbf{318} &  549 & 933  \\
NGC 6366          & 0.57 & 2.34 &   47 &   34 & $0.015 \pm 0.003$ &     0 &  1.88 & \textbf{6.58} &  11.8 &  22.2 &    0 &   61 & \textbf{267} &  514 & 1027 \\
NGC 6397          & 0.61 & 2.18 &   89 &   78 & $0.068 \pm 0.004$ &     0 &     0 & \textbf{1.5} &  4.26 &  8.86 &    0 &    0 & \textbf{ 81} &  230 & 474  \\
NGC 6535          & 1.99 &  4.8 &   20 &   14 & $0.062 \pm 0.015$ &     0 &  0.21 & \textbf{2.61} &  7.07 &  13.2 &    0 &    8 & \textbf{122} &  334 & 627  \\
NGC 6541          & 1.62 & 1.42 &  277 &  438 & $0.081 \pm 0.020$ &     0 &  0.58 & \textbf{3.32} &  8.01 &  13.8 &    0 &   23 & \textbf{155} &  391 & 689  \\
NGC 6584          & 2.45 & 1.12 &   91 &  204 & $0.038 \pm 0.018$ &     0 &  1.82 & \textbf{6.08} &  10.9 &  16.0 &    0 &   67 & \textbf{255} &  491 & 757  \\
NGC 6624          & 1.99 & 1.02 &   73 &  169 & $0.147 \pm 0.051$ &     0 &   0.0 & \textbf{0.25} &  0.76 &  1.72 &    0 &    0 & \textbf{  8} &   27 & 60   \\
NGC 6637 (M69)    & 1.99 &  -   & 200* &  195 & $0.061 \pm 0.026$ &     0 &  6.29 & \textbf{14.6} &  20.9 &  30.8 &    0 &  239 & \textbf{577} &  866 & 1364 \\
NGC 6652          & 3.48 &  -   &  96* &   79 & $0.090 \pm 0.032$ &     0 &   0.4 & \textbf{2.61} &  6.63 &  11.6 &    0 &   14 & \textbf{112} &  292 & 525  \\
NGC 6656 (M22)    & 0.52 & 2.15 &  416 &  430 & $0.026 \pm 0.002$ &     0 &  1.13 & \textbf{6.61} &  13.3 &  37.8 &    0 &   39 & \textbf{303} &  624 & 1956 \\
NGC 6681 (M70)    & 2.45 &  2.0 &  113 &  121 & $0.080 \pm 0.026$ &     0 &  1.48 & \textbf{5.94} &  11.6 &  19.2 &    0 &   58 & \textbf{256} &  534 & 898  \\
NGC 6715 (M54)    & 2.25 & 2.04 & 1410 & 1680 & $0.104 \pm 0.009$ &     0 &  0.21 & \textbf{2.38} &  6.35 &  11.8 &    0 &    9 & \textbf{117} &  313 & 587  \\
NGC 6717 (Pal9)   & 2.45 &  -   &  22* &   31 & $0.064 \pm 0.020$ &     0 &  0.13 & \textbf{2.17} &  5.81 &  11.0 &    0 &    6 & \textbf{106} &  288 & 546  \\
NGC 6723          & 1.15 & 1.77 &  157 &  232 & $0.012 \pm 0.005$ &  0.48 &  7.73 & \textbf{19.0} &  29.3 &  60.1 &    0 &  313 & \textbf{792} & 1297 & 2884 \\
NGC 6752          & 0.91 & 2.17 &  239 &  211 & $0.069 \pm 0.013$ &     0 &  0.06 & \textbf{2.09} &  5.73 &  11.3 &    0 &    3 & \textbf{107} &  297 & 583  \\
NGC 6779 (M56)    & 1.62 & 1.58 &  281 &  157 & $0.029 \pm 0.007$ &  0.36 &   4.2 & \textbf{9.03} &  13.9 &  18.3 &    0 &  153 & \textbf{380} &  610 & 829  \\
NGC 6809 (M55)    & 0.61 & 2.38 &  188 &  182 & $0.010 \pm 0.002$ &  1.59 &  7.78 & \textbf{18.3} &  41.7 &  72.9 &    0 &  247 & \textbf{823} & 1946 & 3739 \\
NGC 6838 (M71)    & 1.00 & 2.76 &   49 &   30 & $0.015 \pm 0.004$ &  0.87 &   6.2 & \textbf{17.3} &  31.3 &  61.6 &    0 &  243 & \textbf{740} & 1400 & 2946 \\
NGC 6934          & 2.45 & 1.76 &  117 &  163 & $0.060 \pm 0.024$ &     0 &   1.3 & \textbf{4.98} &  9.62 &  15.1 &    0 &   47 & \textbf{199} &  414 & 661  \\
NGC 6981 (M72)    & 1.70 &  -   &  63* &  112 & $0.005 \pm 0.004$ &  4.29 &  13.4 & \textbf{21.4} &  34.6 &  48.2 &  152 &  529 & \textbf{908} & 1473 & 2768 \\
NGC 7078 (M15)    & 1.70 & 1.15 &  453 &  811 & $0.111 \pm 0.009$ &     0 &  0.27 & \textbf{2.55} &  6.71 &  12.4 &    0 &   12 & \textbf{126} &  336 & 620  \\
NGC 7089 (M2)     & 1.70 & 1.62 &  582 &  700 & $0.109 \pm 0.012$ &     0 &  0.23 & \textbf{2.55} &  6.79 &  12.5 &    0 &   10 & \textbf{124} &  334 & 624  \\
NGC 7099 (M30)    & 1.70 & 1.85 &  133 &  163 & $0.081 \pm 0.017$ &     0 &  0.02 & \textbf{1.94} &  5.31 &  10.5 &    0 &    1 & \textbf{ 98} &  269 & 537  \\
\enddata
\tablecomments{For each cluster (column 1), the applied radial limit from the observed data is listed in column 2. The cluster mass-to-light ratios computed in \citet{Baumgardt2018} are listed in column 3. Deviations form the standard mass-to-light ratio of 2 help to explain the differences between the total mass estimates in column 4-5. The mass estimates in column 4 are taken from Table 2 of \citet{Baumgardt2018} -- except when marked by an asterisk, in which case they are not listed by the aforementioned source and are instead taken from Table 2 of \citet{Mandushev1991}. Meanwhile, cluster masses in column 5 are computed from the integrated V-band magnitudes in \citet[][2010 edition]{Harris1996}, assuming a uniform mass-to-light ratio of 2. Both sets of masses can be used to multiply the tabulated confidence intervals on $\nbh/\ncluster$ (columns 7-11) and $\mbh/\mcluster$ (columns 12-16) to obtain $\nbh$ and $\mbh$, respectively, as in \autoref{T:bh_dr_Baumgardt}. Note that the tabulated $\nbh/\ncluster$ and $\mbh/\mcluster$ predictions (columns 7-16) are based on $\Deltarfifty$, specifically $\Deltarfifty$ between \texttt{Pop1}, \texttt{Pop2}, and \texttt{Pop3} (see text). Values based on $\Deltaa$, being nearly identical (see \autoref{f:6}), are presented in \autoref{T:raw_results_dA} of the Appendix. Finally, to compare mass segregation between clusters, the $\Deltarfifty^{24}$ values used in \autoref{f:4} (with the uniform choice of $\rlim=0.52\rhl$) are reported in column 6. These $\Delta$ values have Gaussian-shaped uncertainties imposed during the incompleteness correction. As with columns 7-16, the $\Deltaa$ version of column 6 is also reported in \autoref{T:raw_results_dA}.}
\label{T:raw_results_dr}
\end{deluxetable*}

\floattable
\begin{deluxetable*}{l|rrrrr|rrrrr}
\tabletypesize{\scriptsize}
\tablecolumns{11}
\tablewidth{0pt}
\tablecaption{Predicted Number and Total Mass of Retained BHs ($\Deltarfifty$+Baumgardt)}
\tablehead{ & & & & & & & & & & \vspace{-4pt}\\
   \multirow{2}{*}{Cluster} & \multicolumn{5}{c}{$\nbh$} & \multicolumn{5}{c}{$\mbh\ [\msun]$} \vspace{0.07cm} \\
   & $-1\sigma$ & $-2\sigma$ & Mode & $+1\sigma$ & $+2\sigma$ & $-1\sigma$ & $-2\sigma$ & Mode & $+1\sigma$ & $+2\sigma$}
\startdata
NGC 0104 (47Tuc) &   0 &   6 & \textbf{ 43} & 107 & 189 &    0 &  101 & \textbf{  911} &  2353 &  4323 \\
NGC 0288         &   3 &  14 & \textbf{ 26} &  38 &  52 &   45 &  305 & \textbf{  594} &   889 &  1216 \\
NGC 1261         &   4 &  20 & \textbf{ 39} &  60 &  81 &   45 &  416 & \textbf{  845} &  1288 &  1742 \\
NGC 1851         &   0 &   3 & \textbf{ 19} &  47 &  85 &    0 &   57 & \textbf{  420} &  1039 &  1884 \\
NGC 2298         &   0 &   0 & \textbf{  1} &   2 &   3 &    0 &    4 & \textbf{   21} &    47 &    80 \\
NGC 2808         &   0 &  27 & \textbf{ 87} & 153 & 224 &    0 &  416 & \textbf{ 1573} &  3027 &  4682 \\
NGC 3201         &   0 &   7 & \textbf{ 41} &  81 & 187 &    0 &  100 & \textbf{  897} &  1857 &  4865 \\
NGC 4147         &   0 &   0 & \textbf{  2} &   6 &   9 &    0 &    9 & \textbf{   55} &   135 &   235 \\
NGC 4590 (M68)   &   0 &   6 & \textbf{ 17} &  27 &  38 &    0 &  103 & \textbf{  338} &   600 &   888 \\
NGC 4833         &   0 &  22 & \textbf{ 62} & 105 & 209 &    0 &  405 & \textbf{ 1351} &  2376 &  5162 \\
NGC 5024 (M53)   &   0 &  17 & \textbf{ 51} &  87 & 128 &    0 &  296 & \textbf{ 1053} &  1957 &  2991 \\
NGC 5053         &  14 &  20 & \textbf{ 55} &  69 & 104 &  295 &  430 & \textbf{ 1383} &  1727 &  2597 \\
NGC 5272 (M3)    &   0 &   4 & \textbf{ 25} &  64 & 112 &    0 &   71 & \textbf{  587} &  1537 &  2786 \\
NGC 5286         &   0 &   2 & \textbf{ 20} &  54 & 100 &    0 &   40 & \textbf{  493} &  1331 &  2494 \\
NGC 5466         &   5 &  10 & \textbf{ 19} &  39 &  67 &   70 &  197 & \textbf{  423} &   912 &  1740 \\
NGC 5904 (M5)    &   0 &  14 & \textbf{ 45} &  82 & 124 &    0 &  249 & \textbf{  904} &  1722 &  2693 \\
NGC 5927         &  24 &  69 & \textbf{123} & 173 & 273 &  375 & 1320 & \textbf{ 2499} &  3675 &  6110 \\
NGC 5986         &   2 &  43 & \textbf{ 85} & 128 & 199 &    0 &  858 & \textbf{ 1845} &  2875 &  4584 \\
NGC 6093 (M80)   &   0 &   2 & \textbf{ 15} &  38 &  68 &    0 &   45 & \textbf{  356} &   929 &  1673 \\
NGC 6101         &  75 & 104 & \textbf{125} & 190 & 236 & 1748 & 2497 & \textbf{ 3051} &  4976 &  5880 \\
NGC 6144         &   1 &   7 & \textbf{ 13} &  20 &  36 &    0 &  144 & \textbf{  299} &   457 &   855 \\
NGC 6171 (M107)  &   0 &   9 & \textbf{ 23} &  43 &  77 &    0 &  180 & \textbf{  512} &   954 &  1808 \\
NGC 6205 (M13)   &   0 &  61 & \textbf{128} & 196 & 345 &    0 & 1178 & \textbf{ 2786} &  4458 &  8444 \\
NGC 6218 (M12)   &   0 &  11 & \textbf{ 22} &  35 &  65 &    0 &  233 & \textbf{  509} &   822 &  1507 \\
NGC 6254 (M10)   &   0 &  12 & \textbf{ 30} &  49 &  69 &    0 &  206 & \textbf{  622} &  1075 &  1612 \\
NGC 6304         &   0 &  15 & \textbf{ 70} & 102 & 151 &    0 &  227 & \textbf{  726} &  1260 &  1814 \\
NGC 6341 (M92)   &   0 &   3 & \textbf{ 19} &  44 &  75 &    0 &   70 & \textbf{  431} &  1077 &  1876 \\
NGC 6352         &   0 &   5 & \textbf{ 14} &  23 &  39 &    0 &   98 & \textbf{  298} &   515 &   875 \\
NGC 6366         &   0 &   2 & \textbf{  6} &  11 &  21 &    0 &   29 & \textbf{  126} &   243 &   486 \\
NGC 6397         &   0 &   0 & \textbf{  3} &   8 &  16 &    0 &    0 & \textbf{   72} &   204 &   421 \\
NGC 6535         &   0 &   0 & \textbf{  1} &   3 &   5 &    0 &    2 & \textbf{   24} &    67 &   125 \\
NGC 6541         &   0 &   3 & \textbf{ 18} &  44 &  76 &    0 &   64 & \textbf{  429} &  1083 &  1909 \\
NGC 6584         &   0 &   3 & \textbf{ 11} &  20 &  29 &    0 &   61 & \textbf{  231} &   445 &   687 \\
NGC 6624         &   0 &   0 & \textbf{  0} &   1 &   3 &    0 &    0 & \textbf{    6} &    20 &    44 \\
NGC 6637 (M69)   &   0 &  25 & \textbf{ 58} &  84 & 123 &    0 &  478 & \textbf{ 1154} &  1732 &  2728 \\
NGC 6652         &   0 &   1 & \textbf{  5} &  13 &  22 &    0 &   13 & \textbf{  107} &   279 &   501 \\
NGC 6656 (M22)   &   0 &   9 & \textbf{ 55} & 111 & 314 &    0 &  162 & \textbf{ 1260} &  2596 &  8137 \\
NGC 6681 (M70)   &   0 &   3 & \textbf{ 13} &  26 &  43 &    0 &   66 & \textbf{  289} &   603 &  1015 \\
NGC 6715 (M54)   &   0 &   6 & \textbf{ 67} & 179 & 333 &    0 &  127 & \textbf{ 1650} &  4413 &  8277 \\
NGC 6717 (Pal9)  &   0 &   0 & \textbf{  1} &   3 &   5 &    0 &    1 & \textbf{   23} &    63 &   120 \\
NGC 6723         &   2 &  24 & \textbf{ 60} &  92 & 189 &    0 &  491 & \textbf{ 1243} &  2036 &  4528 \\
NGC 6752         &   0 &   0 & \textbf{ 10} &  27 &  54 &    0 &    7 & \textbf{  256} &   710 &  1393 \\
NGC 6779 (M56)   &   2 &  24 & \textbf{ 51} &  78 & 103 &    0 &  430 & \textbf{ 1068} &  1714 &  2329 \\
NGC 6809 (M55)   &   6 &  29 & \textbf{ 69} & 157 & 274 &    0 &  464 & \textbf{ 1547} &  3658 &  7029 \\
NGC 6838 (M71)   &   1 &   6 & \textbf{ 17} &  31 &  60 &    0 &  119 & \textbf{  363} &   687 &  1446 \\
NGC 6934         &   0 &   3 & \textbf{ 12} &  23 &  35 &    0 &   55 & \textbf{  233} &   484 &   773 \\
NGC 6981 (M72)   &   5 &  17 & \textbf{ 27} &  44 &  61 &   96 &  334 & \textbf{  573} &   929 &  1747 \\
NGC 7078 (M15)   &   0 &   2 & \textbf{ 23} &  61 & 112 &    0 &   54 & \textbf{  571} &  1522 &  2809 \\
NGC 7089 (M2)    &   0 &   3 & \textbf{ 30} &  79 & 146 &    0 &   58 & \textbf{  722} &  1944 &  3632 \\
NGC 7099 (M30)   &   0 &   0 & \textbf{  5} &  14 &  28 &    0 &    1 & \textbf{  130} &   358 &   714 \\
\enddata
\tablecomments{Mode and mode-centric confidence intervals ($1\sigma$,$2\sigma$) are presented for $\nbh$ and $\mbh$ in each GC, using the Baumgardt/Mandushev masses in column 4 of \autoref{T:raw_results_dr} to convert from $\nbh/\ncluster$ and $\mbh/\mcluster$. These predictions are based on the mass segregation parameter $\Deltarfifty$. For equivalent predictions based on $\Deltaa$, as well as the Harris masses in column 5 of \autoref{T:raw_results_dr}, see Tables \ref{T:bh_dA_Baumgardt}-\ref{T:bh_dA_Harris} of the Appendix.}
\label{T:bh_dr_Baumgardt}
\end{deluxetable*}

\begin{figure*}
\epsscale{1.177}
\plotone{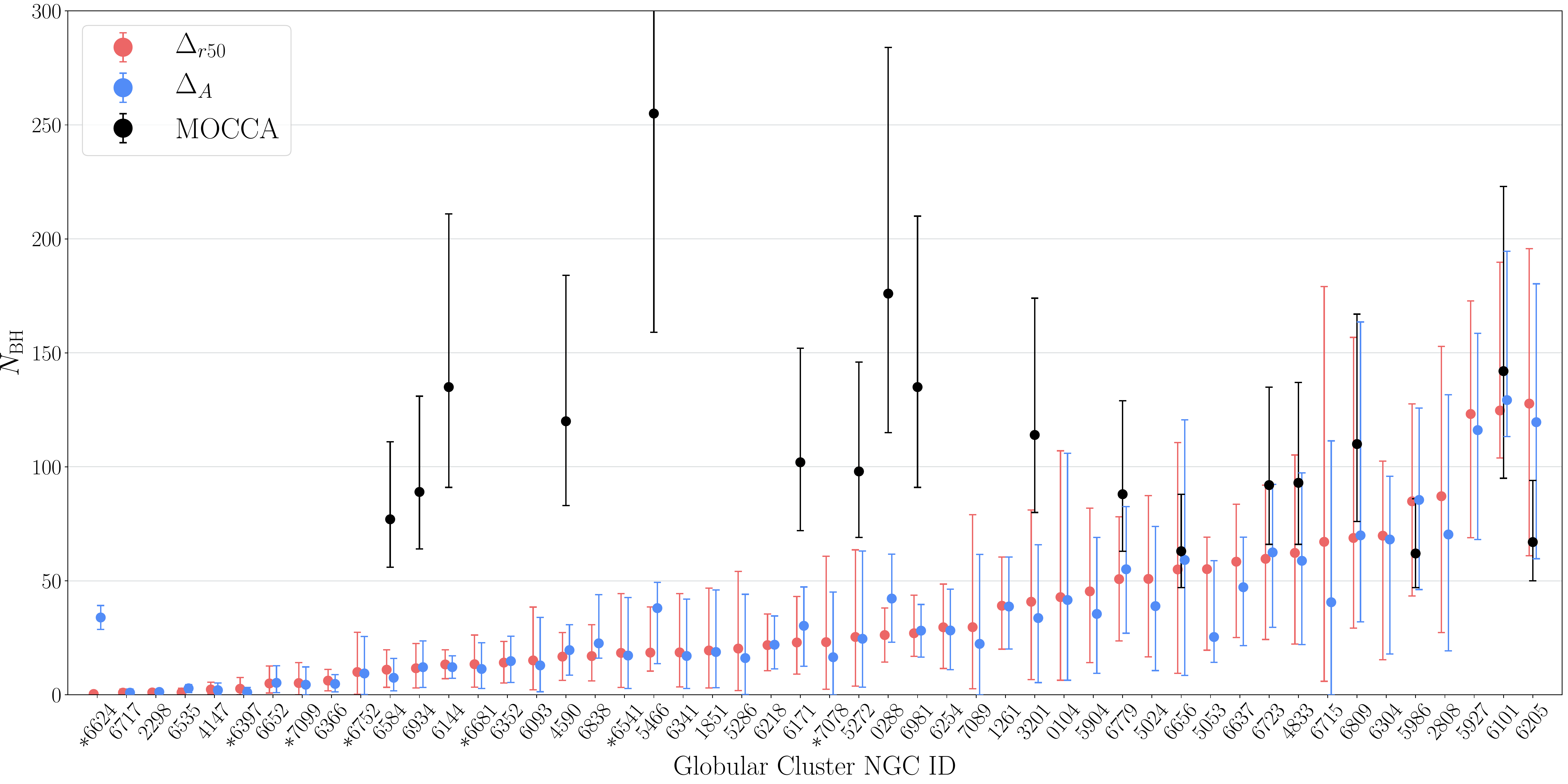}
\plotone{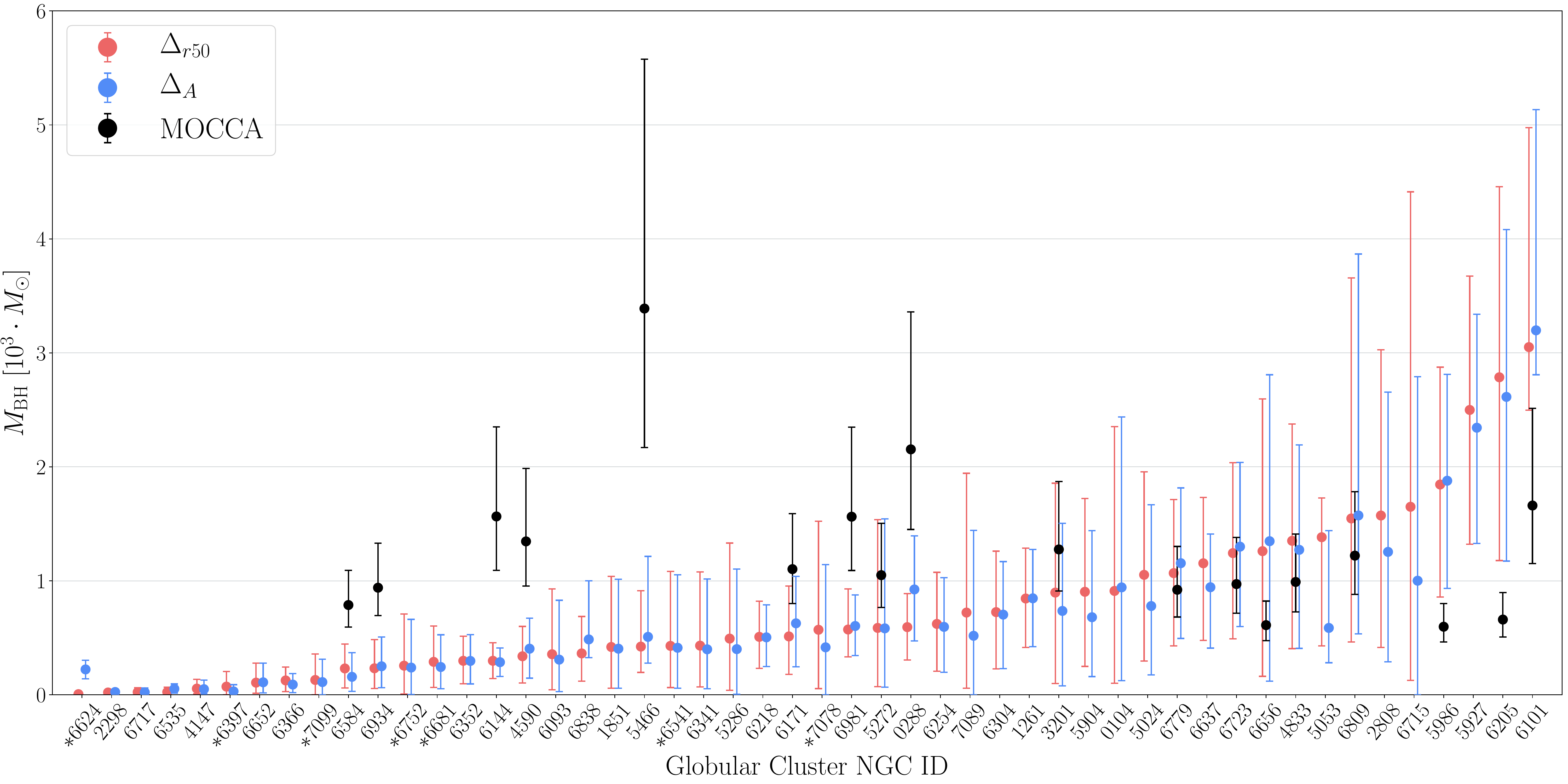}
\caption{Number (panel 1) and total mass (panel 2) of retained BHs for each of the 50 GCs analyzed, sorted in order from lowest to highest $\nbh$ ($\mbh$). Points and uncertainty bars represent the the mode and $1\sigma$ confidence interval, respectively. The results from both mass segregation parameters ($\Deltarfifty$ in red vs $\Deltaa$ in blue) are shown to give strongly consistent predictions. Simply for comparison, the predictions from the MOCCA survey \citep[Table 2 of][]{Askar2018} are shown in black. GCs classified as core-collapsed by \citet{Trager1995} are denoted by an asterisk.}
\label{f:6}
\end{figure*}

To obtain final $\nbh$ and $\mbh$ estimates (not normalized by total cluster mass or star count) we assume an average stellar mass of $0.5\,\msun$ and therefore multiply $\nbh/\ncluster$ ($\mbh/\mcluster$) by twice (once) $\mcluster$. We utilize the total cluster mass estimates in columns 4-5 of \autoref{T:raw_results_dr} based on scaled-up $N$-body simulations \citep{Baumgardt2018,Mandushev1991} as well as values computed from the integrated V-band magnitudes in \citet[][2010 edition]{Harris1996}, assuming a uniform cluster mass-to-light ($M/L$) ratio of two. While the former mass values (henceforth, `Baumgardt/Mandushev') are not purely observational, introducing modeling uncertainties, they \textit{do} account for variation in the cluster $M/L$ ratio, which can differ significantly from the standard value of two in some GCs (see column 3 of \autoref{T:raw_results_dr}). Meanwhile, the latter estimates (henceforth, `Harris') are purely observational but do not account for variation in the $M/L$ ratio. Among the 50 GCs analyzed, the Harris mass values are only about 25\% higher, on average, than those from Baumgardt/Mandushev. However, the difference exceeds a factor of two for a few clusters.

Given the above trade-off between a purely observational approach which does not consider $M/L$ variation, versus a model-based approach which does consider $M/L$ variation but possibly depends on model assumptions, we present our predicted $\nbh$ and $\mbh$ based on both approaches in \autoref{T:bh_dr_Baumgardt} and Tables \ref{T:bh_dA_Baumgardt}-\ref{T:bh_dA_Harris}, leaving it to the readers to decide which estimate is more applicable.\footnote{Note that our direct prediction is the ratio $\nbh/\ncluster$ ($\mbh/\mcluster$) (\autoref{T:raw_results_dr}), hence, it is simple for readers to use any measure of their choice for $\ncluster$ ($\mcluster$) to obtain $\nbh$ ($\mbh$) using our predictions. The above adopted catalogues for $\ncluster$ ($\mcluster$) are simply what we consider the best available examples to use at present.} 
Each table contains the modes, $1\sigma$, and $2\sigma$ CIs on $\nbh$ and $\mbh$ for each of the 50 GCs based on $\Deltaa$ and $\Deltarfifty$, and the two mass estimates mentioned above. Since the differences between the four sets of predictions are generally small compared to the inherent uncertainties in each estimate, we focus our discussion in the rest of the paper only on the $\nbh$ and $\mbh$ predictions in \autoref{T:bh_dr_Baumgardt}, which is based on $\Deltarfifty$ and the Baumgardt/Mandushev masses.

\autoref{f:6} shows the modes and $1\sigma$ CIs for $\nbh$ (top panel) and $\mbh$ (bottom panel), using the Baumgardt/Mandushev masses. Excepting the case of NGC 6624 -- the most mass-segregated cluster in our sample -- the minimal effect of the choice of $\Deltarfifty$ (red) versus $\Deltaa$ (blue) is evident. Furthermore, in the majority of the MWGCs analyzed ($36/50$), observed mass segregation suggests the GC retains a relatively small BH subsystem consisting of fewer than $50$ BHs with a combined mass less than $10^3\,\msun$. 
Nevertheless, we can rule out zero retained BHs at $95\%$ confidence only in 13 MWGCs.
Our survey pinpoints a few MWGCs that are likely to host a large BH subsystem with $\nbh > 80$ ($\mbh > 1,500\,\msun$): NGCs 2808, 5927, 5986, 6101, and 6205.
Interestingly, \citet{ArcaSedda2018} identified the latter three as possibly hosting large BHSs.  Even earlier, \citet{Peuten2016} identified NGC 6101 to contain a BHS. We identify NGCs 2808 and 5927 as two new candidates to host large BHSs.

%

\section{The Role of Black holes in the Evolution of Core Radius} \label{S:corecollapse}

%

As described in \autoref{intro}, the evolution of a cluster, especially the cluster's core structure, is tied to stellar-mass BH dynamics. When a large number of BHs are retained, the energy generated through BH burning is sufficient to delay the onset of core collapse. As the number of retained BHs decreases, so too does the cluster's core radius ($r_c$), until ultimately, the core collapses completely. This connection between core structure and $\nbh$ has been pointed out by a number of recent theoretical studies \citep[e.g.,][]{Mackey2007,Mackey2008,Chatterjee2017a,Kremer2018a,Kremer2020,Askar2018}.

\begin{figure*}[htb!]
\epsscale{1.177}
\plotone{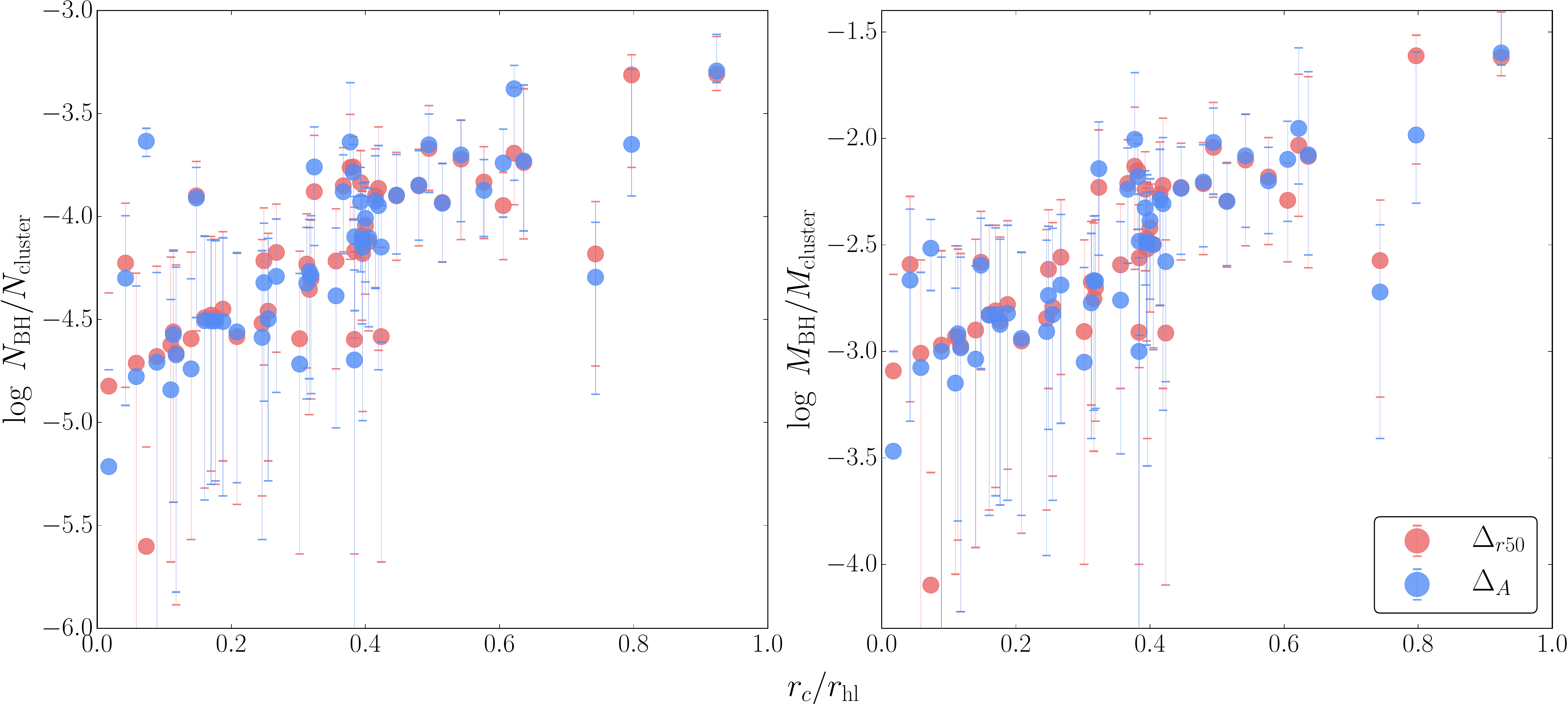}
\caption{Number (left panel) and total mass (right panel) of retained BHs -- normalized by the total number and mass of cluster stars, respectively -- vs $r_c/\rhl$ for each of the 50 GCs analyzed. Points and uncertainty bars represent the modes and $1\sigma$ confidence intervals, respectively. As in \autoref{f:6}, the results from both mass segregation parameters are shown ($\Deltarfifty$ in red, $\Deltaa$ in blue). The correlation shown here between $\nbh/\ncluster$ ($\mbh/\mcluster$) and $r_c/\rhl$ can be attributed to BH burning, as described in the text.}
\label{f:7}
\end{figure*}

In \autoref{f:7}, we show $r_c/r_{\rm{hl}}$ \citep[taken from][2010 edition]{Harris1996} versus our predicted $\nbh/\ncluster$ (left panel) and $\mbh/\mcluster$ (right panel) for the 50 MWGCs we have analyzed. The uncertainty bars denote $1\sigma$ confidence intervals, red and blue denote predictions using $\Deltarfifty$ and $\Deltaa$, respectively.
The figure shows that $\nbh/\ncluster$ and $\mbh/\mcluster$ correlate prominently with $r_c/\rhl$; we predict high $\nbh/\ncluster$ ($\mbh/\mcluster$) in MWGCs with large $r_c/\rhl$. This validates the connection between core evolution and BH dynamics suggested in theoretical studies. For additional detail on this point, see especially Figure 3 of \citet{Kremer2020}, which shows how the number (total mass) and cumulative radial distributions of BHs vary with core radius across our models. In general, nearly $100\%$ of BHs retained in our models at late times reside within the cluster's core radius.

%
\section{Comparison with Prior Results} \label{S:comparison}
%

Our primary finding is that many MWGCs contain non-negligible BH populations at present. However, the number and total mass of BHs in these populations are less than predicted in previous analyses (with some exceptions). We here discuss our predictions in relation to those prior findings, both from models and XRB observations. We especially examine the discrepancy between our results and those of \cite{Askar2018}, currently the only other set of $\nbh$ and $\mbh$ predictions across multiple GCs.

Before comparing with results from other groups, however, it is first important to check for consistency between our new, fully developed $\nbh$ predictions and our trial predictions presented in W1 for the MWGCs 47\,Tuc, M\,10 and M\,22. As discussed in the preceding sections, the three primary differences between the old and new methods are the choice of populations used to quantify mass segregation, the details of the KDE formulation, and the estimated masses of observed GCs. Looking only at $\Deltarfifty$ and re-scaling the old results using the new GC masses (\autoref{T:raw_results_dr}), the new (old) $\nbh$ predictions for these respective clusters are: $43^{+64}_{-37}$ ($21^{+57}_{-17}$), $30^{+19}_{-18}$ ($44^{+26}_{-22}$), and $55^{+56}_{-46}$ ($70^{+72}_{-48}$). The shifts (up for 47\,Tuc, down for M\,10 and M\,22) are well within the $1\sigma$ uncertainty of the predictions. As expected, the new methodology yields results consistent with W1.

\begin{figure*}
\gridline{\fig{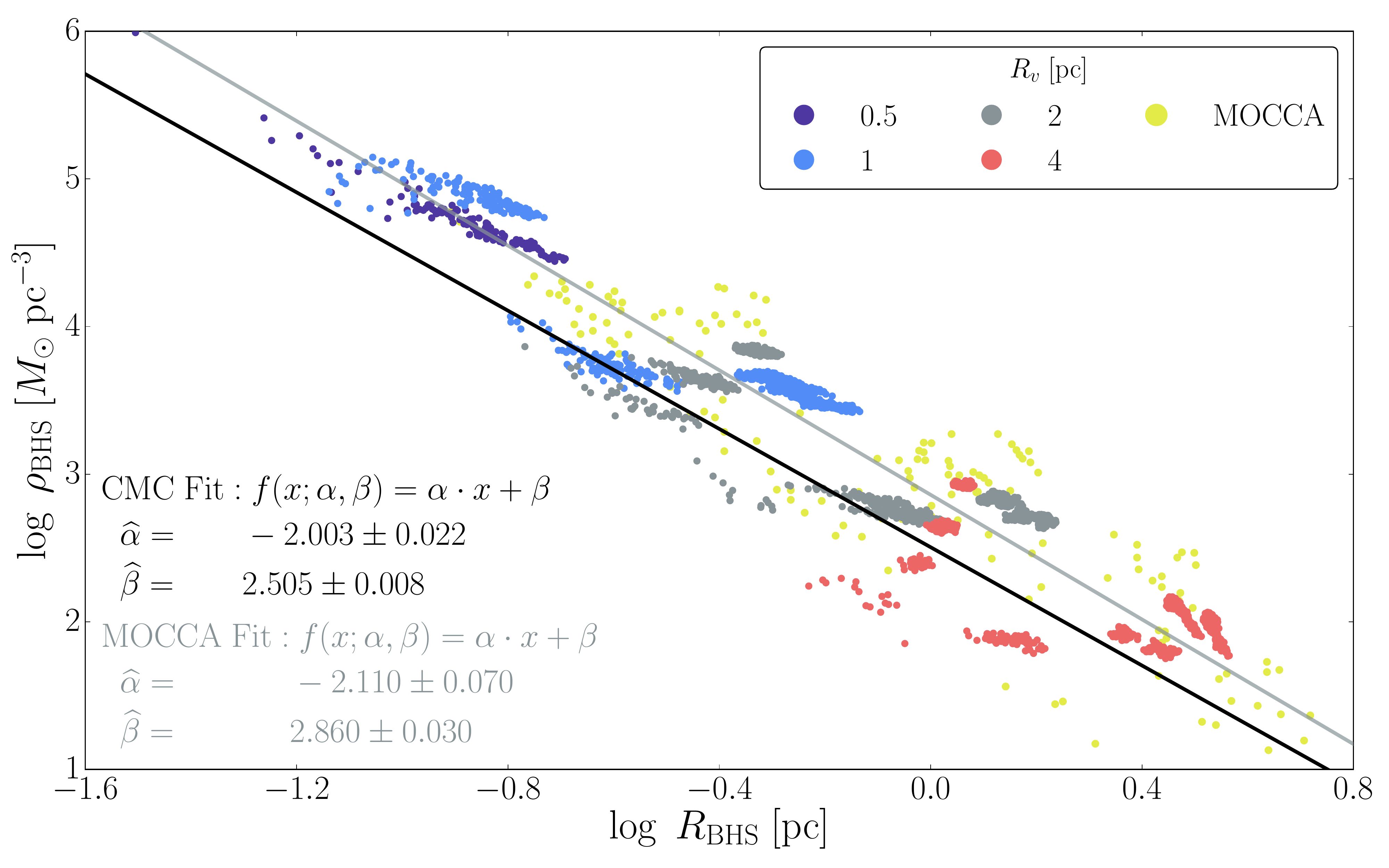}{0.48\textwidth}{(a) BHS radius vs. mass density of BHs in BHS}
          \fig{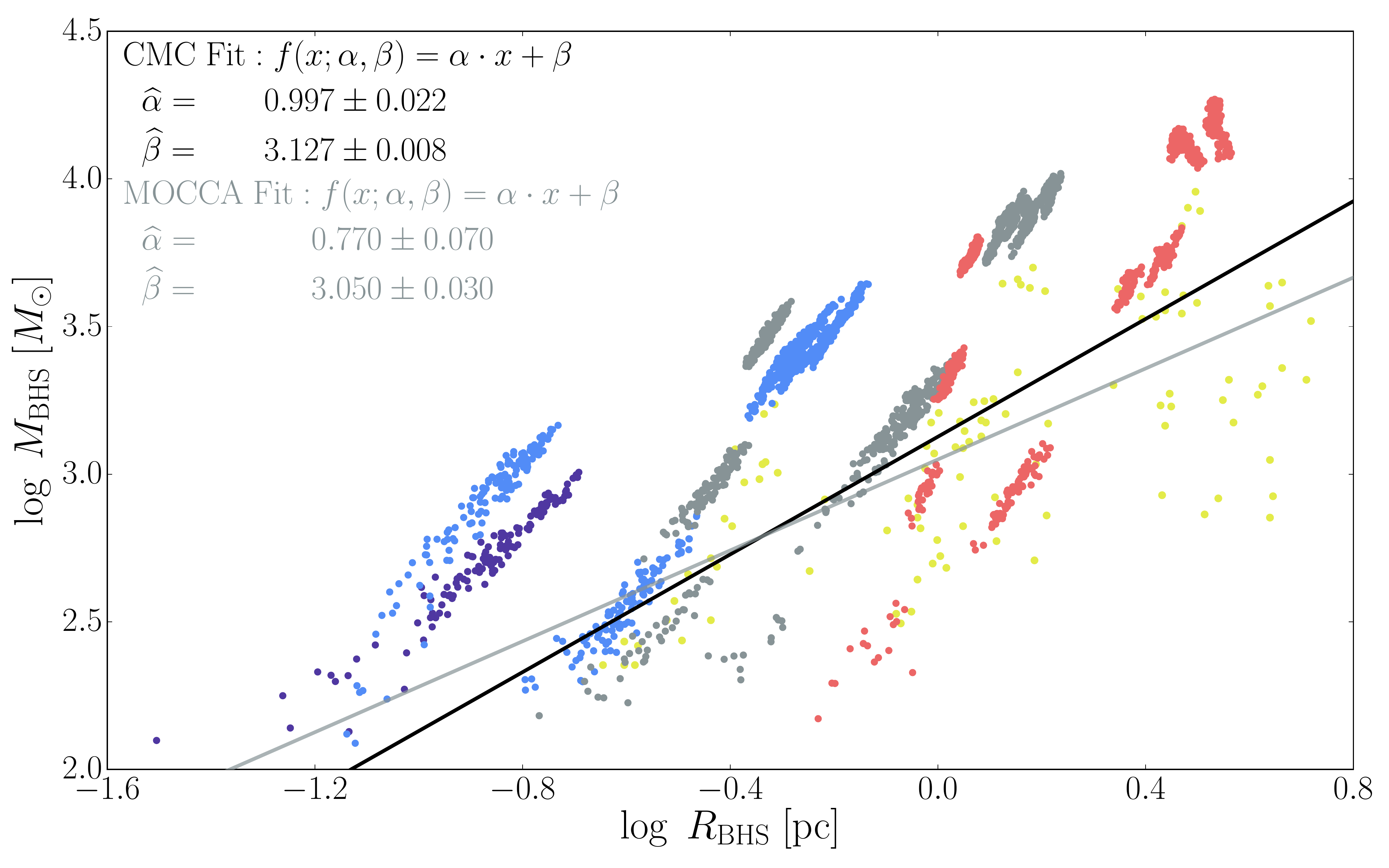}{0.48\textwidth}{(b) BHS radius vs. total mass of BHs in BHS}}
\gridline{\fig{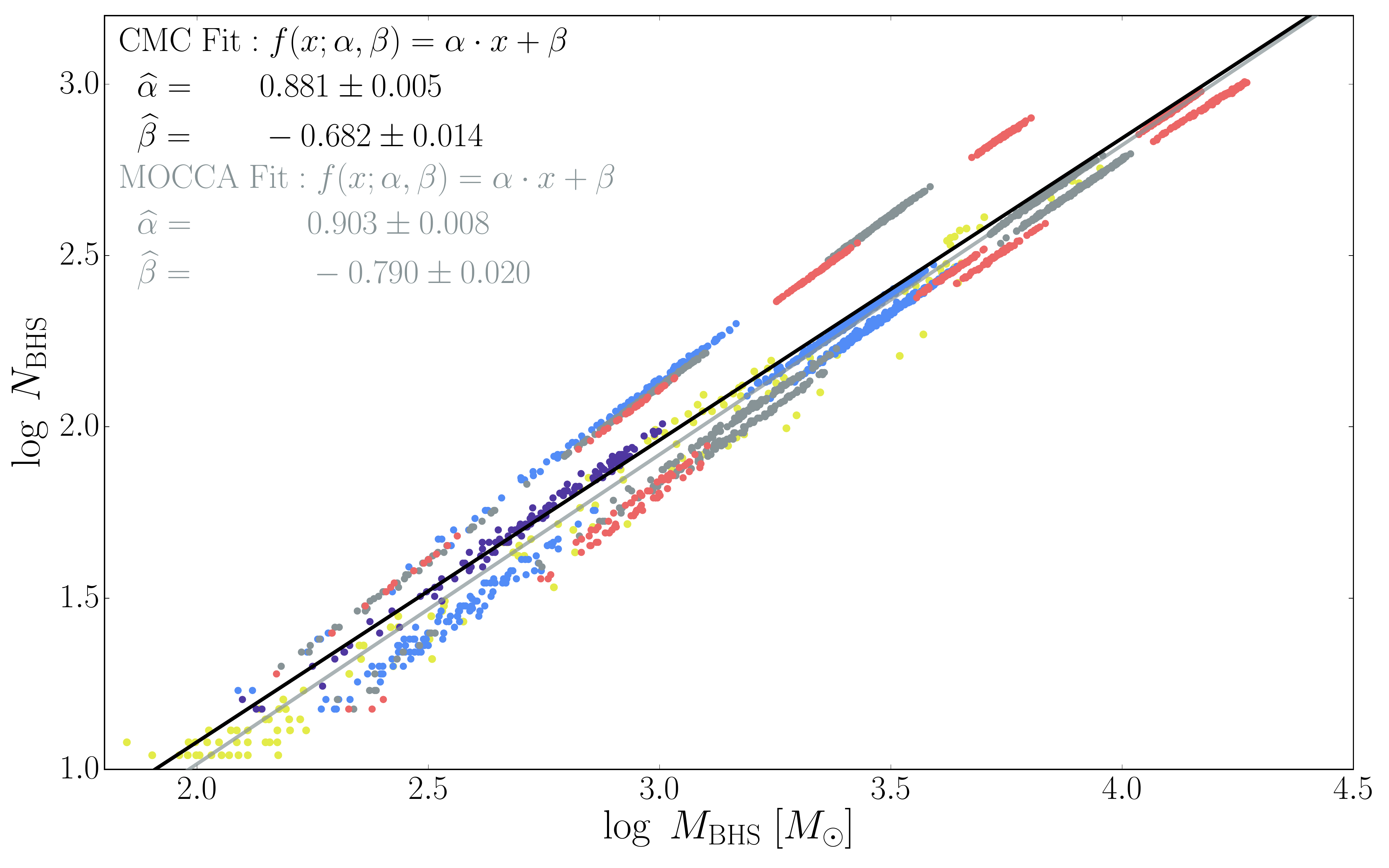}{0.48\textwidth}{(c) Total mass of BHs in BHS vs. number of BHs in BHS}
          \fig{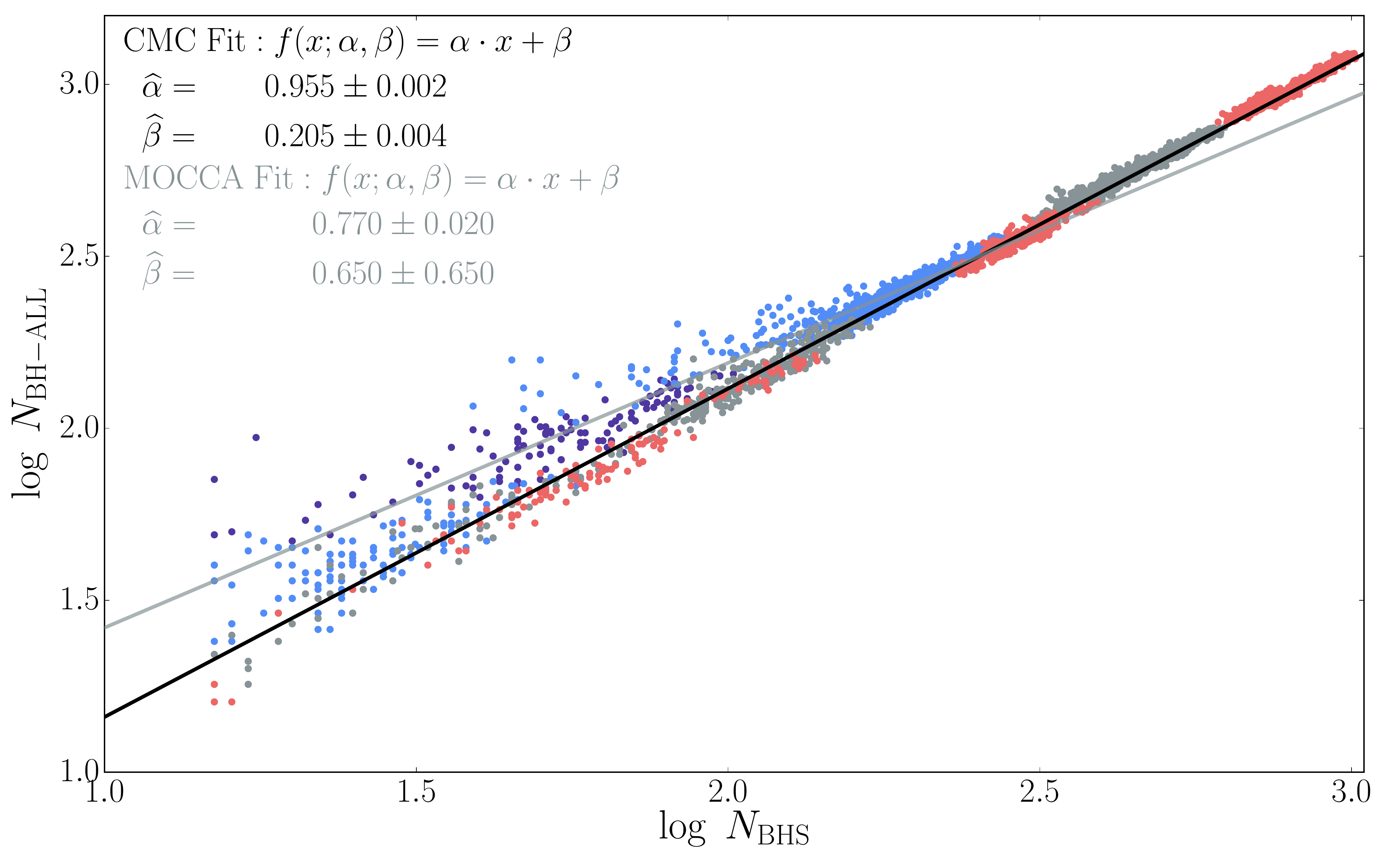}{0.48\textwidth}{(d) Number of BHs in BHS vs. total number of BHs in GC}}
\gridline{\fig{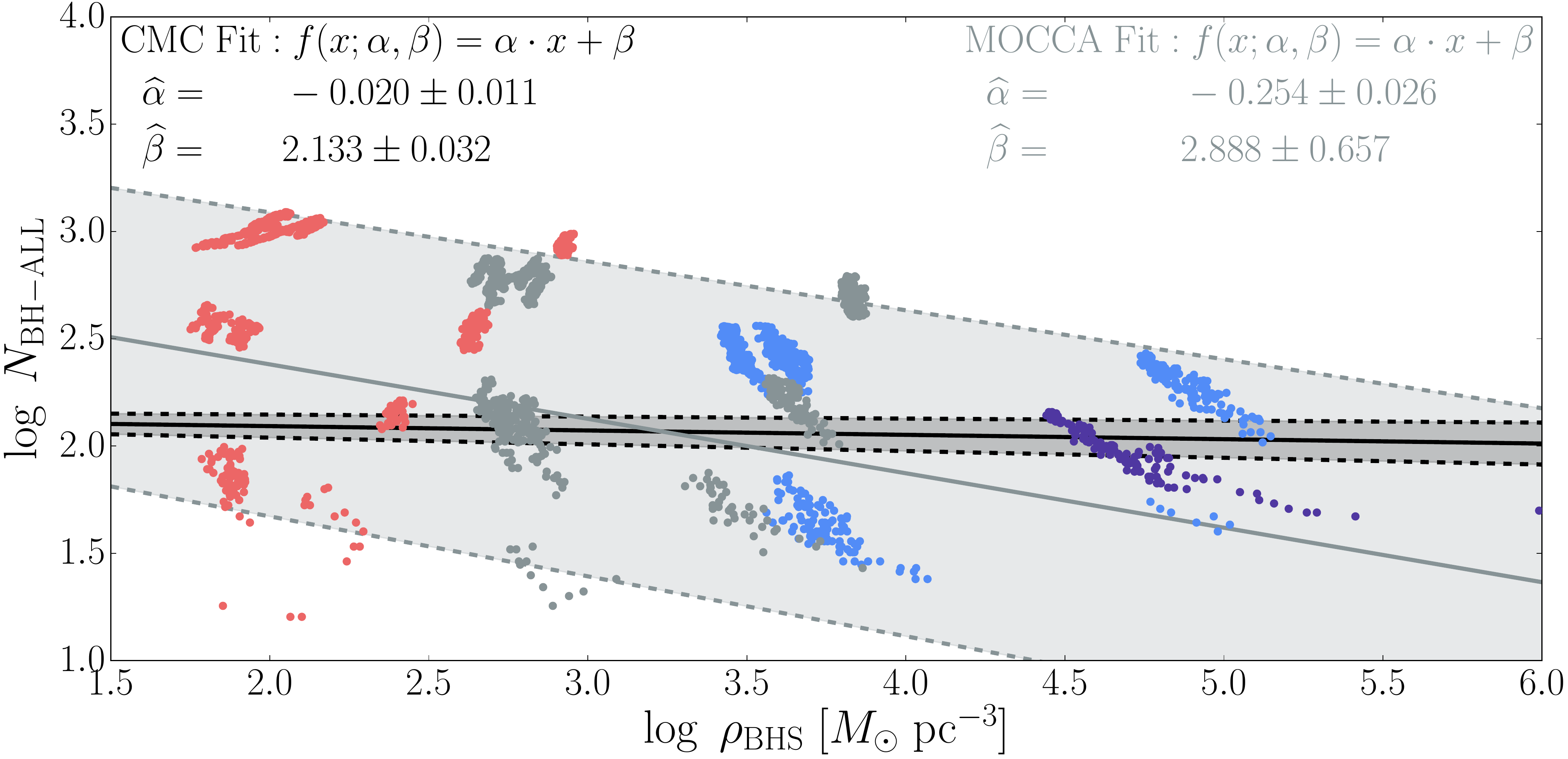}{0.605\textwidth}{(e) Mass density of BHs in BHS vs. total number of BHs in GC}}
\caption{Comparing trends between properties of a cluster's black hole subsystem (BHS) established by the MOCCA Survey group \citep{ArcaSedda2018} to the same trends in our \texttt{CMC} models. Yellow dots \citep[from Figure 3 of][]{ArcaSedda2018} indicate MOCCA models at $12\,\gyr$, with corresponding power law fits \citep[gray lines; from Table A1 of][]{Askar2018}. All other dots, colored by $r_v$ (see legend), correspond to CMC snapshots, also with power law fits (black lines, weighting simulations equally). All parameters are defined in the text. The final panel (e) shows how weak the overall correlation is (between mass density of BHs in the BHS and the total number of BHs in GC) when these two parameters are plotted directly against one another, rather than chaining together the correlations (a-d) like \citet{Askar2018}. Indeed, the direct power-law fit to the \texttt{CMC} models (e; black band) indicates very weak correlation (under $2\sigma$ confidence), while chaining the intermediate fits together, propagating the individual uncertainties, results in spurious correlation (e; gray band).}
\label{f:8}
\end{figure*}

\subsection{Comparison to MOCCA} \label{S:MOCCA}
Concurrently with the publication of W1, the creators of the \textit{MOCCA-Survey Database I} -- another large set of Monte Carlo cluster models similar to those produced by \texttt{CMC} -- developed and continue to use an alternate probe of the BH content in GCs \citep{ArcaSedda2018,Askar2018,ArcaSedda2019}. Both their methodology (discussed below) and results are quite different from our mass segregation approach. We predict lower $\nbh$ in 16 of the 18 MWGCs that the MOCCA team shortlisted as likely BHS hosts. In particularly striking examples, we rule out more than $70$ BHs ($\mbh>1800$) to 95\% confidence in NGCs 0288 and 5466, both of which were predicted to have over $170$ BHs ($\mbh>2000$) in the MOCCA survey. Given these discrepancies, it is essential to more deeply examine the methodology behind the MOCCA results and the benefits or drawbacks relative to our own methods.

First, \cite{ArcaSedda2018} find a set of scaling relations between key properties of the BH subsystem that mass-segregates to the core of a GC. Specifically, they define $R_{\rm{BHS}}$ as the cluster-centric distance within which half of the total mass is in BHs (the other half of the mass is contained in stars). BHs within distance $R_{\rm{BHS}}$ from the cluster center count as members of the subsystem, which then typically contains around 60\% of the total number of BHs in the models. The authors correlate $R_{\rm{BHS}}$ with number ($N_{\rm{BHS}}$) and total mass ($M_{\rm{BHS}}$) of BHs in the subsystem, and anti-correlate these three quantities with the associated BH mass density $\rho_{\rm{BHS}} = M_{\rm{BHS}}/R_{\rm{BHS}}^3$. Finally, they establish a tight model correlation between $\rho_{\rm{BHS}}$ and GC average surface luminosity $L/r_{\rm{hl,obs}}^2$, which they apply in a companion paper \citep{Askar2018} to short-list 29 MWGCs with sizable BH subsystems, using observed V-band magnitudes and half-light radii from the Harris catalog \citep[2010 edition]{Harris1996}. The authors utilize a similar method to identify MWGCs that potentially host an IMBH \citep{ArcaSedda2019}.

Applying the above definitions to our own model set results in similar correlations, but a closer examination reveals several issues. The most critical concern is statistical. Whereas we use \textit{non-parametric} KDEs to directly relate our observables ($\Delta^{23}$, $\Delta^{34}$) to $\nbh/\ncluster$ and $\mbh/\mcluster$, \cite{Askar2018} indirectly chain 5 separate correlations together, each with their own assumed parametric form, to relate their observable ($L_V/\rhl^2$) to $\nbh$. Specifically, linear curve-fits in log-log scale are applied in each of the 5 steps along the following chain: $L_V/\rhl^2$ to $\rho_{\rm{BHS}}$ to $R_{\rm{BHS}}$ to $M_{\rm{BHS}}$ to $N_{\rm{BHS}}$ to $\nbh$ (i.e. $N_{\rm{BH-ALL}}$, the total number of BHs in the cluster). The latter four of these power-law relations are shown in \autoref{f:8} (a-d) for both our own model set and the MOCCA data.

Crucially, all curve fits inherently assume the parametric form used to fit the data is a true representation of the underlying statistical distribution and ignores scatter around the parametric form. They are therefore inherently biased towards the particular form used. Chaining curve fits further amplifies potential bias; the more `links' in the chain, the more any deviations from a perfect fit to the underlying distribution conspire to bias the correlation. 

This distorting effect is easily seen by plotting the observable \textit{directly} against the dependent variable of interest. Skipping the first link in the chain ($L_V/\rhl^2$ to $\rho_{\rm{BHS}}$), we plot $\rho_{\rm{BHS}}$ vs $\nbh$ in the bottom panel of \autoref{f:8}, bypassing the intermediate variables $R_{\rm{BHS}}$, $M_{\rm{BHS}}$, and $N_{\rm{BHS}}$. Applying a single, unchained least-squares fit to the \texttt{CMC} data (black band), it is evident that there is only a very weak anti-correlation between $\rho_{\rm{BHS}}$ and $\nbh$. This contradicts the analysis by \citet{Askar2018}; chaining together their step-wise correlations from the upper panels results in a much larger anti-correlation between $\rho_{\rm{BHS}}$ and $\nbh$ (gray band). Propagating uncertainty, the $1\sigma$ confidence interval (gray band) on predicted $\nbh$ spans nearly 2 orders of magnitude for any given value of $\rho_{\rm{BHS}}$. The inflated uncertainty partially mitigates, but does not eliminate, the underlying bias introduced from chaining multiple parametric fits.


As mentioned in Section 2.6 of \cite{Askar2018}, models without $\nbh > 15$ at $12\,\gyr$ were excluded from their analysis. Of these excluded models, most incorporated high BH natal kicks, but $\sim 40$ utilized the standard mass fallback prescription \citep{Belczynski2002} and had \textit{similar values of the observable} ($L_V/\rhl^2$) to the $163$ included models, despite having significantly \textit{fewer} BHs ($\nbh < 15$). This indicates that $L_V/\rhl^2$ may not actually be a strong predictor of the BH content in GCs, supporting our findings in the bottom panel of \autoref{f:6}. At the very least, excluding the 20\% of models with lowest $\nbh$ would naturally cause the MOCCA team to over-predict $\nbh$ and $\mbh$, partially explaining why our analysis generally yields lower predictions.

\subsection{Possible Sources of Uncertainties in Our Predictions} \label{S:uncertainties}
Although our analysis based on the \texttt{CMC} catalog benefits from non-parametric statistical methods that are generally less prone to bias than parametric methods -- especially the MOCCA team's chaining technique -- the models from the MOCCA database do present their own advantages. Whereas the models in the \texttt{CMC} Catalog all start with a primordial binary fraction $f_b = 5\%$, expected to be typical of the MWGCs, the MOCCA models initialize with a variety of binary fractions, ranging from $f_b = 5\%$ all the way up to $95\%$ \citep{Askar2017a}. Though we found in W1 that shifting $f_b$ between $5\%$ and $10\%$ had no distinguishable impact on the anti-correlation between $\Delta$ and $\nbh$, it is conceivable that higher binary fractions could impact the correlation. This possibility could favor the MOCCA team's $\nbh$ ($\mbh$) predictions for any GC where the true $f_b$ is much greater than $5\%$.

Other potential improvements to both our own analysis and that of the MOCCA team may include consideration of primordial mass segregation and non-standard initial mass functions. Although the former is expected to have little effect after several relaxation times, it may accelerate the dynamical evolution of BHs. In addition, primordial mass segregation may change the BH mass function via collisions and accretions onto the BHs \citep{Kremer2020}, thus indirectly affecting mass segregation.

Meanwhile, the choice of IMF, especially the slope for the high-mass stars, has a dramatic impact on cluster evolution in general \citep[e.g.,][]{Chatterjee2017a}. 
However, the effects of non-standard IMFs on the $\Delta$--$\nbh$ ($\mbh$) anti-correlation may be more subtle. Any IMF that simply changes the total number and average mass of the BHs themselves may not significantly affect the predicted numbers -- fewer or less-massive BHs would still lead to higher $\Delta$. This robustness is in fact a unique feature of our approach. Perhaps more likely to affect the anti-correlation are IMFs which fundamentally alter the relative importance of other dynamically influential populations, such as neutron stars. Additionally, if we assume that the non-standard IMF varies from cluster-to-cluster, then the inherent spread in the $\Delta$--$\nbh/\ncluster$ anti-correlation in our models will likely increase (\autoref{f:2}), swelling the uncertainty on $\nbh$ ($\mbh$) predicted in real MWGCs. Further studying the late-time effects of both primordial mass segregation and non-standard IMFs may therefore prove illuminating in future work.

Ultimately, we find that quantities like $\Delta$ that parameterize mass segregation are a more reliable statistical predictor of the total mass and number of BHs inside a GC than the $L_V/\rhl^2$-BHS correlations used in the MOCCA survey \citep{ArcaSedda2018,Askar2018,ArcaSedda2019}, This, at least, is true independently of any other advantages or disadvantages (discussed above) associated with the specific model sets the analysis must rely upon.

Constituting the largest  sample  of  GCs  for  which BH populations have been reported to-date, our analysis suggests while some BH retention is common to many GCs, fewer are retained -- generally less than $50$ -- than has typically been suggested previously. In addition to this general point, we discuss findings of particular interest for specific MWGCs in the following subsections.

\subsection{47\,Tuc} \label{S:47Tuc}
As one of the nearest and therefore most well-studied GCs, 47\,Tuc (NGC 0104) is an important cluster for benchmarking. The cluster's mass of around $10^6\,\msun$ (\autoref{T:raw_results_dr}) is near the maximum of our model space at $13\,\gyr$, but its Galactocentric distance and metallicity are well within the model bounds \citep[][2010 edition]{Harris1996}. In W1, our $\nbh$ predictions for 47\,Tuc were limited by a dearth of models with high mass segregation. Now, without such a limitation, we predict the cluster retains more BHs, around 40 totalling 900 $\msun$. This new estimate is still well within $1\sigma$ of the estimate in W1. Our current estimate is also consistent to $1\sigma$ with a contemporary study to specifically model 47\,Tuc that predicts a relatively small BHS in the cluster \citep{Henault-Brunet2020}.
Note however, at 95\% confidence, we can neither exclude zero BHs nor a large population of up to $\sim200$ BHs totalling 4,300 $\msun$ in 47\,Tuc.

\subsection{NGC 2808} \label{S:NGC2808}
\cite{Lutzgendorf2012a} previously found five high-velocity giants in the core of NGC 2808 and suggested their extreme velocities could have resulted from close encounters with a stellar-mass BH or IMBH, but most likely the former with a mass of about 10 $\msun$. In follow-up analysis with Monte Carlo 3-body scattering experiments, they further solidified this hypothesis and constrained the maximum mass of the BH to be no more than $10^4\,\msun$ \citep{Lutzgendorf2012b}. These prior findings fit well with our observation that NGC 2808 is one of the least mass-segregated clusters in the MW.  The low observed $\Delta$ leads to our prediction that NGC 2808 contains around 90 BHs totalling 1,500 $\msun$ in mass (\autoref{T:bh_dr_Baumgardt}). Taken together, these lines of evidence strongly suggest that NGC 2808 presently retains a robust central BH population.

\subsection{NGC 3201} \label{S:NGC3201}
Recently, \citet{Giesers2018} reported a stellar-mass BH in the cluster NGC 3201. They inferred the  BH's presence from the large radial velocity variations ($\sim 100$ km/s) of an apparently lone main sequence star, thereby presumed to be orbiting a compact remnant. This detection -- along with two more recent ones \citep{Giesers2019} -- made NGC 3201 the fifth MWGC known to harbor a stellar-mass BH candidate. Shortly thereafter, \cite{Kremer2018b} used \texttt{CMC} to model the cluster, reporting that it likely retains $>200$ stellar-mass BHs at present, an estimate that was revised down to $\nbh = 120 \pm 10$ in a follow-up using updated BH formation physics \citep{Kremer2019a}. This revised prediction is in line with the MOCCA team's estimate, $\nbh = 114^{+60}_{-35}$ \citep{Askar2018}, but mass segregation predicts an even lower number: $\nbh = 41^{+40}_{-34}$.

\subsection{NGC 6101} \label{S:NGC6101}
Of the 50 MWGCs surveyed, NGC 6101 is the least mass-segregated and is by far the best candidate in which to find a large number of BHs. To 95\% confidence, we estimate it contains $75-236$ BHs with a combined mass of $1,750-5,900\,\msun$. Most likely, it contains $\sim 125$ BHs totalling $\sim 3,000\,\msun$. This conclusion is supported by a growing body of evidence from other sources. \cite{Dalessandro2015} were the first to draw attention to this GC's unusually low mass segregation, finding no evidence for the phenomenon based on three different measures: the radial distribution of blue stragglers, that of MS binaries, and the luminosity function. Following this finding, \cite{Peuten2016} and \cite{Webb2017} explored the anti-correlation between $\nbh$ and mass segregation in $N$-body simulations to demonstrate that the cluster may contain a large population of BHs. \cite{BaumgardtSollima2017} disputed these suggestions because their estimates of NGC 6101's mass-function slope indicated mass segregation after all. However, given that this rebuttal relies on the same ACS dataset as applied in this study and because their results similarly suggest that NGC 6101 has one of the lowest levels of mass segregation among MWGCs, we find no contradiction to our conclusions; NGC 6101 is very likely to host a robust population of stellar-mass BHs. This determination is further supported by the findings of \cite{Askar2018} discussed above.



\subsection{NGC 6535} \label{S:NGC6535}
NGC 6535 is unusual in that it's relatively old but has a high mass-to-light ratio in the range 5 \citep{Baumgardt2018} to 11 \citep{Zaritsky2014}. \cite{Halford2015} found that its observed mass-function has a positive slope -- indicating a high loss-rate of low-mass stars and making its high $M/L$ ratio even more puzzling. Given NGC 6535's small Galactocentric distance of $3.9$ kpc \citep[2010 edition]{Harris1996}, it is likely that increased tidal stripping of low-mass stars near the Galactic center is responsible for the positive mass-function slope. However, \cite{Halford2015} found no evidence that clusters near the Galactic center with similarly top-heavy mass functions had artificially inflated mass estimates, raising the possibility that some dark mass may be responsible for NGC 6535's high $M/L$ ratio. Recently, \cite{Askar2017b} demonstrated that $N$-body simulations of clusters containing an IMBH or BHS were able to fit the photometric and kinematic properties of NGC 6535, but later concluded the cluster contains neither a significant BHS nor an IMBH \citep{Askar2018,ArcaSedda2019}. Since we rule out more than $130\,\msun$ of BHs in NGC 6535 to 95\% confidence, the mystery of the apparently missing mass in this cluster remains an open question.

\subsection{NGC 6624} \label{S:NGC6624}
\cite{Perera2017} reported the possible presence of an IMBH in NGC 6624 based on timing observations of a millisecond pulsar near the projected cluster center. Their timing analysis indicated the presence of an IMBH with mass in the range 7,500 to 10,000 $\msun$, even up to 60,000 $\msun$. This finding was disputed by \cite{Gieles2018}, who demonstrated that dynamical models without an IMBH produce maximum accelerations at the pulsar's position comparable to its observed line-of-sight acceleration. Recently, \cite{Baumgardt2019} similarly found that their $N$-body models without an IMBH could provide excellent fits to the observed velocity dispersion and surface brightness profiles (VDPs and SDPs) in NGC 6624. Their cluster models with an IMBH indicated that an IMBH in NGC 6624 with mass $>$1,000 $\msun$ was incompatible with the cluster's observed VDP and SBP. Meanwhile, based on data from HST and ATCA, \cite{Tremou2018} found that all radio emissions observed from NGC 6624 are consistent with being from a known ultra-compact X-ray binary in the cluster's core. Their radio observations place a $3\sigma$ upper limit on the cluster's possible IMBH mass of 1,550 $\msun$. Although we have yet to explore how much difference an IMBH has on quenching $\Delta$ compared to a BHS, our results support the latter three studies; we find to 95\% confidence that there are no more than $\sim400\,\msun$ of BHs in NGC 6624 (using Baumgardt's cluster mass, otherwise $<\sim900\,\msun$ of BHs using Harris' cluster mass). Indeed, NGC 6624 is the most mass-segregated cluster in our sample, suggesting that it may in fact be one of the MWGCs \textit{least} likely to host an IMBH or significant BHS.

\subsection{M\,54} \label{S:M54}
Thought to be a MWGC for over two centuries, the cluster M\,54 (NGC 6715) is now known to be coincident with the center of the Sagittarius Dwarf Galaxy \citep[e.g.,][]{Monaco2005}, perhaps even as the galaxy's original nucleus \citep{Layden2000}. While M\,54's metallicity is well-covered by our model parameter space, its effective Galactocentric distance is unreliable because our models assume a MW-like potential for tidal boundary calculations. Its approximate mass is also at the extreme upper end of the model space (\autoref{T:raw_results_dr}). Therefore, with some reservation, despite M\,54's highly mass-segregated present state, we predict a significant number of BHs remain in the cluster at present, with $67^{+112}_{-61}$ BHs totalling around $1,650^{+2,763}_{-1,523}$ $\msun$. This prediction is consistent with the $3\sigma$ upper limit on a single accreting IMBH of $<3,000\,\msun$ imposed by VLA radio observations \citep{Tremou2018}.

\subsection{NGC 6723} \label{S:NGC6723}
NGC 6723 is listed as possibly core-collapsed in the Harris catalog \citep[][2010 edition]{Harris1996}, an identification which would appear at odds with our relatively high prediction for $\nbh$ and $\mbh$ in this GC (\autoref{f:6}, \autoref{T:bh_dr_Baumgardt}). However, Table 2 of \citet{Trager1995}, the purported source material for the Harris catalog's core-collapsed classifications, lists NGC 6723 as non-core-collapsed. Their surface brightness profile for the cluster also shows a flat core density, further contradicting the Harris classification. We speculate that the Harris catalog may have accidentally swapped the core-collapse classifications between NGCs 6723 and 6717 (Palomar 9), which appear consecutively in Table 2 of \citep{Trager1995}. For this reason, we do not mark this MWGC as core-collapsed in \autoref{f:6}.
%
\section{Summary \& Discussion} \label{S:summary}
%

\subsection{Summary}

We have presented a statistically robust method that uses mass segregation between easily observable stellar populations to determine the number of BHs in a cluster. Our process can be implemented for any observed MWGC and carefully accounts for potential sources of bias between models and observations, including field-of-view limits, projection effects, and observational incompleteness. Due to the expansive grid of realistic cluster models used, the process also accounts for many uncertainties on cluster initial conditions. We briefly summarize our key findings below.

\begin{enumerate}
    \item We demonstrated that, overall, the \texttt{CMC} Cluster Catalog models yield mass segregation ($\Delta$) values which closely match the observed distribution in $\Delta$ among real MWGCs (see \autoref{f:4}). This provides strong evidence that our models capture the state of mass segregation in realistic MWGCs, complementing the results of \citet{Kremer2020}.
    \item By using $\Delta$ as a predictive parameter, we have constrained the total number and mass in stellar-mass BHs contained in more MWGCs, 50 total, than any prior studies.
    \item We find that 35 of the 50 GCs studied retain more than 20 BHs at present and 8 retain more than 80 BHs. These predictions indicate that present-day BH retention is common to many MWGCs, though to a lesser extent than suggested in competing analyses, \citep[e.g.,][]{Askar2018}.
    \item Specifically, we have identified NGCs 2808, 5927, 5986, 6101, and 6205 to contain especially large BH populations, each with total BH mass exceeding $10^3\,\msun$. These clusters may serve as ideal observational targets for BH candidate searches.
    \item We also explored in detail the advantages and disadvantages of our statistical methods compared to other similar analyses in the literature.
\end{enumerate}

\subsection{Discussion and Future Work}


Here, we predict smaller BH populations in a few GCs compared to our previous analyses which also utilized \texttt{CMC} models \citep[e.g.,][]{Kremer2019a}. The exact number of BHs is highly uncertain (indeed, this is reflected by the uncertainty bars in \autoref{f:6} and all the tables). Hence, discrepancy between these results and those of our previous work -- which implemented entirely different methods based on fitting surface brightness and velocity dispersion profiles to predict $\nbh$ -- is unsurprising. Critically, as shown in \autoref{f:7}, the overall connection between cluster core evolution and BH dynamics put forward in previous work \citep{Mackey2008,Kremer2018b,Kremer2019a} is confirmed. This further validates the significant role BHs play in GC evolution.

There are a couple of more speculative conclusions hinted at by our results which are worth mentioning briefly, but require additional study. First, it is tempting to extrapolate our predictions of total BH mass in GCs to place upper limits on the masses of possible intermediate-mass black holes (IMBHs) in those clusters. Indeed, $N$-body simulations have shown that an IMBH of mass $>1\%$ of its host GC's overall mass should significantly quench mass segregation -- even among only visible giants and MS stars \citep[e.g.,][]{Gill2008,Pasquato2016}. The generally significant mass segregation we measure in the 50 GCs studied -- representative of the MW as a whole -- therefore suggests that IMBHs with mass $>$1,000 $\msun$ are rare in MWGCs. However, firmer constraints would require testing beyond the scope of this study, specifically on how similar the dynamical impact of a single IMBH is to that of a stellar-mass BH population with identical total mass. Is it a one-to-one relation, or does a, for example, $1,000\,\msun$ IMBH perhaps have a much weaker effect on mass segregation than a population of a hundred $10\,\msun$ BHs? For now, the prospect of IMBHs in GCs is still best analyzed through direct observations in the X-ray and radio bands, as well as via the accelerations of luminous stars within the IMBH's `influence radius,' but further study may be able to extend our constraints on stellar-mass BH populations to IMBHs in GCs.

Second, it has been suggested that clusters were born already mass-segregated to a degree, a property termed `primordial' mass segregation \citep[e.g.,][]{Baumgardt2008}. Our models assume clusters have no primordial mass segregation. Hence, the close match between $\Delta$ in our models and the $\Delta$ distribution observed in the MWGCs (see \autoref{f:4}) demonstrates that our models do not need to start off with some degree of mass segregation to match real clusters. This finding could suggest that primordial mass segregation is minimal or non-existent in the MWGCs, but such a conclusion is tenuous since primordial mass segregation is likely to be washed out at the present day after many relaxation times. Further consideration of the late-time effects of primordial mass segregation on presently observable $\Delta$ is necessary to make any further conclusions on this matter.

Finally, although mass segregation has been shown here to be a strong indicator of BH populations in clusters, recent analyses have shown that many other observables, including millisecond pulsars \citep{Ye2019}, blue stragglers \citep{Kremer2020}, and cluster surface brightness and velocity dispersion profiles \citep[e.g.,][]{Mackey2008,Kremer2018b}, may also correlate with BH dynamics and thus may also serve as indicators of retained BH numbers. In order to pin down more precisely the true number of BHs retained in specific clusters, all of these observables should be leveraged in tandem. We intend to pursue such analysis further in future works.


\acknowledgements
We thank Mario Spera for detailed comments on the manuscript and Claire Ye, Nicholas Rui, and Giacomo Fragione for useful discussions throughout the preparation of this work.
This work was supported by NSF Grant AST-1716762 and through the computational resources and staff contributions provided for the {\tt Quest} high performance computing facility at Northwestern University. {\tt Quest} is jointly supported by the Office of the Provost, the Office for Research, and Northwestern University Information Technology. SC acknowledges support from NASA through Chandra Award Number TM5-16004X issued by the Chandra X-ray Observatory Center
(operated by the Smithsonian Astrophysical Observatory for and on behalf of NASA under contract NAS8-03060), and from the Department of Atomic Energy, Government of India, under project no. 12-R\&D-TFR-5.02-0200. 
K.K.\ acknowledges support by the National Science Foundation Graduate Research Fellowship Program under Grant No. DGE-1324585.

\software{\texttt{CMC} \citep{Joshi2000, Joshi2001, Fregeau2003, Fregeau2007, Chatterjee2010, Umbreit2012, Pattabiraman2013, Chatterjee2013, Morscher2015, Rodriguez2016, Rodriguez2018, Kremer2020}.}

\bibliography{bhsurvey}

\appendix
\setcounter{table}{0}
\renewcommand{\thetable}{A\arabic{table}}

\vspace{-15pt}
\floattable
\begin{deluxetable*}{lccccc|rrrrr|rrrrr}[b!]
\centerwidetable
\tabletypesize{\scriptsize}
\tablecolumns{16}
\tablewidth{0pt}
\tablecaption{Cluster Properties and Raw Computational Results Based on $\Deltaa$}
\tablehead{ & & & & & & & & & & & & & & & \vspace{-4pt}\\
   \multirow{2}{*}{Cluster} & \multirow{2}{*}{$\displaystyle{\frac{\rlim}{\rhl}}$} & \multirow{2}{*}{$\displaystyle{\frac{\mcluster}{\lcluster}}$} & \multicolumn{2}{c}{$\mcluster/(10^3\cdot\msun)$} & $\Deltaa^{24}$ & \multicolumn{5}{c}{$(\nbh/\ncluster)\cdot 10^5$} & \multicolumn{5}{c}{$(\mbh/\mcluster)\cdot 10^5$} \vspace{0.00cm} \\
   & & & Baumgardt & Harris & $\pmsigma$ & $-2\sigma$ & $-1\sigma$ & Mode & $+1\sigma$ & $+2\sigma$ &
   $-2\sigma$ & $-1\sigma$ & Mode & $+1\sigma$ & $+2\sigma$}
\startdata
NGC 0104 (47Tuc)  & 0.55 & 1.77 &  779 & 1000 & $0.052 \pm 0.006$ &     0 &  0.41 & \textbf{2.67} &   6.8 &  11.9 &    0 &   16 & \textbf{121} &  313 & 558  \\
NGC 0288          & 0.77 & 2.39 &  116 &   87 & $0.008 \pm 0.001$ &  2.24 &  9.93 & \textbf{18.2} &  26.6 &  46.9 &   44 &  408 & \textbf{797} & 1202 & 2219 \\
NGC 1261          & 2.45 & 2.12 &  167 &  225 & $0.020 \pm 0.010$ &  1.25 &  6.02 & \textbf{11.6} &  18.1 &  23.6 &   40 &  253 & \textbf{507} &  764 & 1004 \\
NGC 1851          & 3.48 & 2.02 &  302 &  367 & $0.093 \pm 0.032$ &     0 &  0.52 & \textbf{3.11} &  7.61 &  13.2 &    0 &   19 & \textbf{134} &  336 & 590  \\
NGC 2298          & 1.70 & 0.46 &   12 &   57 & $0.022 \pm 0.003$ &     0 &  1.63 & \textbf{5.41} &  9.68 &  14.5 &    0 &   53 & \textbf{215} &  432 & 696  \\
NGC 2808          & 2.25 & 1.64 &  742 &  975 & $0.067 \pm 0.009$ &     0 &   1.3 & \textbf{4.74} &  8.87 &  13.4 &    0 &   39 & \textbf{169} &  358 & 578  \\
NGC 3201          & 0.57 &  2.4 &  149 &  163 & $0.012 \pm 0.001$ &     0 &   1.8 & \textbf{11.3} &  22.1 &  54.9 &    0 &   53 & \textbf{494} & 1010 & 2810 \\
NGC 4147          & 3.48 & 1.51 &   33 &   50 & $0.043 \pm 0.009$ &     0 &  0.44 & \textbf{3.09} &  7.88 &  14.0 &    0 &   20 & \textbf{151} &  391 & 698  \\
NGC 4590 (M68)    & 1.15 & 2.02 &  123 &  152 & $0.014 \pm 0.002$ &  0.12 &  3.49 & \textbf{7.97} &  12.5 &  16.8 &    0 &  119 & \textbf{329} &  547 & 772  \\
NGC 4833          & 0.73 & 0.84 &  247 &  317 & $0.011 \pm 0.001$ &     0 &  4.47 & \textbf{11.9} &  19.7 &  38.5 &    0 &  165 & \textbf{515} &  888 & 1864 \\
NGC 5024 (M53)    & 1.29 & 1.59 &  380 &  521 & $0.034 \pm 0.006$ &     0 &   1.4 & \textbf{5.12} &  9.72 &  15.0 &    0 &   46 & \textbf{205} &  439 & 714  \\
NGC 5053          & 0.68 & 1.66 &   57 &   87 & $0.005 \pm 0.001$ &   7.3 &  12.6 & \textbf{22.4} &  52.0 &  83.6 &  291 &  497 & \textbf{1037} & 2546 & 4276 \\
NGC 5272 (M3)     & 0.77 & 1.56 &  394 &  610 & $0.036 \pm 0.006$ &     0 &  0.42 & \textbf{3.12} &  8.01 &  14.3 &    0 &   17 & \textbf{148} &  392 & 716  \\
NGC 5286          & 2.25 & 1.41 &  401 &  536 & $0.099 \pm 0.012$ &     0 &  0.02 & \textbf{2.01} &   5.5 &  10.9 &    0 &    1 & \textbf{100} &  275 & 548  \\
NGC 5466          & 0.77 & 1.13 &   46 &  106 & $0.004 \pm 0.001$ &  9.64 &  15.0 & \textbf{41.7} &  54.1 &  83.2 &  400 &  610 & \textbf{1115} & 2664 & 4261 \\
NGC 5904 (M5)     & 1.00 & 1.52 &  372 &  572 & $0.033 \pm 0.005$ &     0 &  1.27 & \textbf{4.77} &  9.28 &  14.5 &    0 &   43 & \textbf{183} &  387 & 626  \\
NGC 5927          & 1.62 & 2.61 &  354 &  228 & $0.016 \pm 0.006$ &  2.62 &  9.62 & \textbf{16.4} &  22.4 &  32.9 &   78 &  375 & \textbf{662} &  943 & 1447 \\
NGC 5986          & 1.70 & 2.45 &  301 &  406 & $0.019 \pm 0.011$ &  0.96 &  7.67 & \textbf{14.2} &  20.9 &  29.6 &    4 &  310 & \textbf{624} &  934 & 1314 \\
NGC 6093 (M80)    & 2.89 & 1.43 &  249 &  335 & $0.111 \pm 0.012$ &     0 &  0.27 & \textbf{2.59} &  6.82 &  12.5 &    0 &   11 & \textbf{124} &  333 & 618  \\
NGC 6101          & 1.62 &  3.0 &  127 &  102 & $0.003 \pm 0.002$ &  35.8 &  44.6 & \textbf{50.9} &  76.6 &  94.7 & 1665 & 2211 & \textbf{2518} & 4043 & 4693 \\
NGC 6144          & 1.00 & 0.54 &   45 &   94 & $0.011 \pm 0.002$ &  1.28 &  8.01 & \textbf{13.4} &  18.9 &  25.9 &   30 &  357 & \textbf{632} &  909 & 1253 \\
NGC 6171 (M107)   & 1.00 & 2.16 &   87 &  121 & $0.015 \pm 0.002$ &  1.18 &  7.22 & \textbf{17.4} &  27.2 &  45.7 &   13 &  283 & \textbf{721} & 1195 & 2106 \\
NGC 6205 (M13)    & 1.00 & 2.61 &  453 &  450 & $0.013 \pm 0.004$ &     0 &  6.59 & \textbf{13.2} &  19.9 &  32.7 &    0 &  259 & \textbf{577} &  901 & 1553 \\
NGC 6218 (M12)    & 0.99 & 1.27 &   87 &  144 & $0.011 \pm 0.002$ &     0 &  6.58 & \textbf{12.7} &  20.0 &  35.6 &    0 &  286 & \textbf{583} &  912 & 1523 \\
NGC 6254 (M10)    & 0.89 & 1.94 &  184 &  168 & $0.018 \pm 0.002$ &     0 &  3.01 & \textbf{7.67} &  12.6 &  17.7 &    0 &  107 & \textbf{324} &  559 & 827  \\
NGC 6304          & 1.29 & 1.37 &  277 &  142 & $0.050 \pm 0.015$ &     0 &  3.23 & \textbf{12.3} &  17.3 &  24.1 &    0 &   83 & \textbf{254} &  422 & 591  \\
NGC 6341 (M92)    & 1.70 & 1.81 &  268 &  329 & $0.066 \pm 0.015$ &     0 &  0.52 & \textbf{3.18} &  7.84 &  13.7 &    0 &   20 & \textbf{149} &  379 & 681  \\
NGC 6352          & 0.85 & 2.47 &   94 &   66 & $0.017 \pm 0.002$ &     0 &  2.91 & \textbf{7.88} &  13.7 &  21.1 &    0 &  102 & \textbf{317} &  563 & 924  \\
NGC 6366          & 0.57 & 2.34 &   47 &   34 & $0.015 \pm 0.001$ &     0 &  1.37 & \textbf{5.07} &  9.38 &  16.7 &    0 &   39 & \textbf{190} &  393 & 785  \\
NGC 6397          & 0.61 & 2.18 &   89 &   78 & $0.059 \pm 0.003$ &     0 &     0 & \textbf{0.61} &   1.8 &  4.06 &    0 &    0 & \textbf{ 34} &  100 & 226  \\
NGC 6535          & 1.99 &  4.8 &   20 &   14 & $0.037 \pm 0.006$ &     0 &  2.45 & \textbf{7.11} &  11.3 &  15.4 &    0 &   72 & \textbf{264} &  478 & 713  \\
NGC 6541          & 1.62 & 1.42 &  277 &  438 & $0.068 \pm 0.013$ &     0 &   0.5 & \textbf{3.11} &  7.71 &  13.5 &    0 &   21 & \textbf{149} &  380 & 677  \\
NGC 6584          & 2.45 & 1.12 &   91 &  204 & $0.025 \pm 0.011$ &     0 &  0.94 & \textbf{4.12} &   8.8 &  14.2 &    0 &   33 & \textbf{174} &  407 & 690  \\
NGC 6624          & 1.99 & 1.02 &   73 &  169 & $0.125 \pm 0.038$ &   0.7 &  19.6 & \textbf{23.2} &  26.8 &  31.1 &   73 &  193 & \textbf{305} &  416 & 525  \\
NGC 6637 (M69)    & 1.99 &  -   & 200* &  195 & $0.046 \pm 0.017$ &   0.3 &  5.39 & \textbf{11.8} &  17.3 &  22.7 &    0 &  205 & \textbf{472} &  705 & 951  \\
NGC 6652          & 3.48 &  -   &  96* &   79 & $0.077 \pm 0.023$ &     0 &  0.51 & \textbf{2.75} &  6.68 &  11.3 &    0 &   17 & \textbf{115} &  292 & 511  \\
NGC 6656 (M22)    & 0.52 & 2.15 &  416 &  430 & $0.018 \pm 0.001$ &     0 &  1.02 & \textbf{7.11} &  14.5 &  40.9 &    0 &   29 & \textbf{324} &  675 & 2104 \\
NGC 6681 (M70)    & 2.45 &  2.0 &  113 &  121 & $0.075 \pm 0.016$ &     0 &  1.21 & \textbf{5.02} &  10.1 &  16.3 &    0 &   47 & \textbf{216} &  466 & 770  \\
NGC 6715 (M54)    & 2.25 & 2.04 & 1410 & 1680 & $0.074 \pm 0.007$ &     0 &     0 & \textbf{1.44} &  3.95 &  8.36 &    0 &    0 & \textbf{ 71} &  198 & 420  \\
NGC 6717 (Pal9)   & 2.45 &  -   &  22* &   31 & $0.047 \pm 0.011$ &     0 &  0.15 & \textbf{2.13} &  5.67 &  10.7 &    0 &    6 & \textbf{104} &  277 & 528  \\
NGC 6723          & 1.15 & 1.77 &  157 &  232 & $0.010 \pm 0.003$ &  1.69 &  9.44 & \textbf{19.9} &  29.4 &  59.5 &   34 &  382 & \textbf{828} & 1299 & 2853 \\
NGC 6752          & 0.91 & 2.17 &  239 &  211 & $0.054 \pm 0.007$ &     0 &  0.02 & \textbf{1.96} &  5.35 &  10.6 &    0 &    1 & \textbf{100} &  277 & 548  \\
NGC 6779 (M56)    & 1.62 & 1.58 &  281 &  157 & $0.019 \pm 0.004$ &  0.65 &  4.81 & \textbf{9.8} &  14.7 &  19.4 &    5 &  176 & \textbf{411} &  646 & 865  \\
NGC 6809 (M55)    & 0.61 & 2.38 &  188 &  182 & $0.009 \pm 0.001$ &   1.9 &  8.51 & \textbf{18.6} &  43.5 &  73.8 &   16 &  284 & \textbf{837} & 2057 & 3786 \\
NGC 6838 (M71)    & 1.00 & 2.76 &   49 &   30 & $0.008 \pm 0.002$ &  9.36 &  16.4 & \textbf{23.0} &  44.7 &  69.7 &  365 &  664 & \textbf{990} & 2038 & 3315 \\
NGC 6934          & 2.45 & 1.76 &  117 &  163 & $0.052 \pm 0.016$ &     0 &  1.38 & \textbf{5.18} &  10.1 &  15.7 &    0 &   54 & \textbf{214} &  433 & 681  \\
NGC 6981 (M72)    & 1.70 &  -   &  63* &  112 & $0.003 \pm 0.003$ &  6.02 &  13.1 & \textbf{22.3} &  31.4 &  45.5 &  235 &  547 & \textbf{957} & 1388 & 2103 \\
NGC 7078 (M15)    & 1.70 & 1.15 &  453 &  811 & $0.102 \pm 0.006$ &     0 &     0 & \textbf{1.82} &  4.98 &  10.0 &    0 &    0 & \textbf{ 92} &  252 & 507  \\
NGC 7089 (M2)     & 1.70 & 1.62 &  582 &  700 & $0.101 \pm 0.008$ &     0 &     0 & \textbf{1.92} &  5.29 &  10.6 &    0 &    0 & \textbf{ 89} &  248 & 502  \\
NGC 7099 (M30)    & 1.70 & 1.85 &  133 &  163 & $0.067 \pm 0.011$ &     0 &     0 & \textbf{1.67} &   4.6 &  9.39 &    0 &    0 & \textbf{ 84} &  235 & 481  \\
\enddata
\tablecomments{This table is identical to \autoref{T:raw_results_dr}, except based on $\Deltarfifty$ instead of $\Deltaa$. Specifically, column 6 lists the $\Deltaa^{24}$ values used in \autoref{f:4} (with the uniform choice of $\rlim=0.52\rhl$). Again, these $\Delta$ values have Gaussian-shaped uncertainties imposed during the incompleteness correction. Similarly, the $\nbh/\ncluster$ and $\mbh/\mcluster$ predictions listed in columns 7-16 are based on $\Deltaa$ between \texttt{Pop1}, \texttt{Pop2}, and \texttt{Pop3} (see \autoref{S:nbh_predictions}).}
\label{T:raw_results_dA}
\end{deluxetable*}

\floattable
\begin{deluxetable*}{l|rrrrr|rrrrr}
\tabletypesize{\scriptsize}
\tablecolumns{11}
\tablewidth{0pt}
\tablecaption{Predicted Number and Mass of Retained BHs ($\Deltaa$+Baumgardt)}
\tablehead{ & & & & & & & & & & \vspace{-4pt}\\
   \multirow{2}{*}{Cluster} & \multicolumn{5}{c}{$\nbh$} & \multicolumn{5}{c}{$\mbh\ [\msun]$} \vspace{0.07cm} \\
   & $-1\sigma$ & $-2\sigma$ & Mode & $+1\sigma$ & $+2\sigma$ & $-1\sigma$ & $-2\sigma$ & Mode & $+1\sigma$ & $+2\sigma$}
\startdata
NGC 0104 (47Tuc) &   0 &   6 & \textbf{ 42} & 106 & 185 &    0 &  125 & \textbf{  943} &  2438 &  4347 \\
NGC 0288         &   5 &  23 & \textbf{ 42} &  62 & 109 &   51 &  473 & \textbf{  925} &  1394 &  2574 \\
NGC 1261         &   4 &  20 & \textbf{ 39} &  60 &  79 &   67 &  423 & \textbf{  847} &  1276 &  1677 \\
NGC 1851         &   0 &   3 & \textbf{ 19} &  46 &  80 &    0 &   57 & \textbf{  405} &  1015 &  1782 \\
NGC 2298         &   0 &   0 & \textbf{  1} &   2 &   3 &    0 &    6 & \textbf{   25} &    50 &    81 \\
NGC 2808         &   0 &  19 & \textbf{ 70} & 132 & 199 &    0 &  289 & \textbf{ 1254} &  2656 &  4289 \\
NGC 3201         &   0 &   5 & \textbf{ 34} &  66 & 164 &    0 &   79 & \textbf{  736} &  1505 &  4187 \\
NGC 4147         &   0 &   0 & \textbf{  2} &   5 &   9 &    0 &    7 & \textbf{   50} &   129 &   230 \\
NGC 4590 (M68)   &   0 &   9 & \textbf{ 20} &  31 &  41 &    0 &  146 & \textbf{  405} &   673 &   950 \\
NGC 4833         &   0 &  22 & \textbf{ 59} &  97 & 190 &    0 &  408 & \textbf{ 1272} &  2193 &  4604 \\
NGC 5024 (M53)   &   0 &  11 & \textbf{ 39} &  74 & 114 &    0 &  175 & \textbf{  779} &  1668 &  2713 \\
NGC 5053         &   8 &  14 & \textbf{ 25} &  59 &  95 &  165 &  281 & \textbf{  587} &  1441 &  2420 \\
NGC 5272 (M3)    &   0 &   3 & \textbf{ 25} &  63 & 113 &    0 &   67 & \textbf{  583} &  1544 &  2821 \\
NGC 5286         &   0 &   0 & \textbf{ 16} &  44 &  87 &    0 &    4 & \textbf{  401} &  1103 &  2197 \\
NGC 5466         &   9 &  14 & \textbf{ 38} &  49 &  76 &  182 &  278 & \textbf{  508} &  1215 &  1943 \\
NGC 5904 (M5)    &   0 &   9 & \textbf{ 35} &  69 & 108 &    0 &  160 & \textbf{  681} &  1440 &  2329 \\
NGC 5927         &  19 &  68 & \textbf{116} & 159 & 233 &  276 & 1328 & \textbf{ 2343} &  3338 &  5122 \\
NGC 5986         &   6 &  46 & \textbf{ 85} & 126 & 178 &   12 &  933 & \textbf{ 1878} &  2811 &  3955 \\
NGC 6093 (M80)   &   0 &   1 & \textbf{ 13} &  34 &  62 &    0 &   27 & \textbf{  309} &   829 &  1539 \\
NGC 6101         &  91 & 113 & \textbf{129} & 195 & 241 & 2115 & 2808 & \textbf{ 3198} &  5135 &  5960 \\
NGC 6144         &   1 &   7 & \textbf{ 12} &  17 &  23 &   14 &  162 & \textbf{  286} &   412 &   568 \\
NGC 6171 (M107)  &   2 &  13 & \textbf{ 30} &  47 &  80 &   11 &  246 & \textbf{  627} &  1040 &  1832 \\
NGC 6205 (M13)   &   0 &  60 & \textbf{120} & 180 & 296 &    0 & 1173 & \textbf{ 2614} &  4082 &  7035 \\
NGC 6218 (M12)   &   0 &  11 & \textbf{ 22} &  35 &  62 &    0 &  247 & \textbf{  504} &   789 &  1317 \\
NGC 6254 (M10)   &   0 &  11 & \textbf{ 28} &  46 &  65 &    0 &  197 & \textbf{  596} &  1029 &  1522 \\
NGC 6304         &   0 &  18 & \textbf{ 68} &  96 & 134 &    0 &  230 & \textbf{  704} &  1169 &  1637 \\
NGC 6341 (M92)   &   0 &   3 & \textbf{ 17} &  42 &  73 &    0 &   54 & \textbf{  399} &  1016 &  1825 \\
NGC 6352         &   0 &   5 & \textbf{ 15} &  26 &  40 &    0 &   96 & \textbf{  297} &   528 &   867 \\
NGC 6366         &   0 &   1 & \textbf{  5} &   9 &  16 &    0 &   18 & \textbf{   90} &   186 &   371 \\
NGC 6397         &   0 &   0 & \textbf{  1} &   3 &   7 &    0 &    0 & \textbf{   30} &    89 &   201 \\
NGC 6535         &   0 &   1 & \textbf{  3} &   5 &   6 &    0 &   14 & \textbf{   53} &    96 &   143 \\
NGC 6541         &   0 &   3 & \textbf{ 17} &  43 &  75 &    0 &   58 & \textbf{  413} &  1053 &  1875 \\
NGC 6584         &   0 &   2 & \textbf{  7} &  16 &  26 &    0 &   30 & \textbf{  158} &   369 &   626 \\
NGC 6624         &   1 &  29 & \textbf{ 34} &  39 &  45 &   53 &  141 & \textbf{  223} &   304 &   384 \\
NGC 6637 (M69)   &   1 &  22 & \textbf{ 47} &  69 &  91 &    0 &  410 & \textbf{  944} &  1410 &  1902 \\
NGC 6652         &   0 &   1 & \textbf{  5} &  13 &  22 &    0 &   16 & \textbf{  110} &   279 &   488 \\
NGC 6656 (M22)   &   0 &   8 & \textbf{ 59} & 121 & 340 &    0 &  121 & \textbf{ 1348} &  2808 &  8753 \\
NGC 6681 (M70)   &   0 &   3 & \textbf{ 11} &  23 &  37 &    0 &   53 & \textbf{  244} &   527 &   870 \\
NGC 6715 (M54)   &   0 &   0 & \textbf{ 41} & 111 & 236 &    0 &    0 & \textbf{ 1001} &  2792 &  5922 \\
NGC 6717 (Pal9)  &   0 &   0 & \textbf{  1} &   2 &   5 &    0 &    1 & \textbf{   23} &    61 &   116 \\
NGC 6723         &   5 &  30 & \textbf{ 62} &  92 & 187 &   53 &  600 & \textbf{ 1300} &  2039 &  4479 \\
NGC 6752         &   0 &   0 & \textbf{  9} &  26 &  51 &    0 &    2 & \textbf{  239} &   662 &  1310 \\
NGC 6779 (M56)   &   4 &  27 & \textbf{ 55} &  83 & 109 &   14 &  495 & \textbf{ 1155} &  1815 &  2431 \\
NGC 6809 (M55)   &   7 &  32 & \textbf{ 70} & 164 & 277 &   30 &  534 & \textbf{ 1574} &  3867 &  7118 \\
NGC 6838 (M71)   &   9 &  16 & \textbf{ 23} &  44 &  68 &  179 &  326 & \textbf{  486} &  1001 &  1628 \\
NGC 6934         &   0 &   3 & \textbf{ 12} &  24 &  37 &    0 &   63 & \textbf{  250} &   507 &   797 \\
NGC 6981 (M72)   &   8 &  17 & \textbf{ 28} &  40 &  57 &  148 &  345 & \textbf{  604} &   876 &  1327 \\
NGC 7078 (M15)   &   0 &   0 & \textbf{ 16} &  45 &  91 &    0 &    0 & \textbf{  417} &  1142 &  2297 \\
NGC 7089 (M2)    &   0 &   0 & \textbf{ 22} &  62 & 123 &    0 &    0 & \textbf{  518} &  1443 &  2922 \\
NGC 7099 (M30)   &   0 &   0 & \textbf{  4} &  12 &  25 &    0 &    0 & \textbf{  112} &   313 &   640 \\
\enddata
\tablecomments{Mode and mode-centric confidence intervals ($1\sigma$,$2\sigma$) are presented for $\nbh$ and $\mbh$ in each GC, using the Baumgardt/Mandushev masses in column 4 of \autoref{T:raw_results_dr} to convert from $\nbh/\ncluster$ and $\mbh/\mcluster$. These predictions are based on the mass segregation parameter $\Deltaa$.}
\label{T:bh_dA_Baumgardt}
\end{deluxetable*}

\floattable
\begin{deluxetable*}{l|rrrrr|rrrrr}
\tabletypesize{\scriptsize}
\tablecolumns{11}
\tablewidth{0pt}
\tablecaption{Predicted Number and Mass of Retained BHs ($\Deltarfifty$+Harris)}
\tablehead{ & & & & & & & & & & \vspace{-4pt}\\
   \multirow{2}{*}{Cluster} & \multicolumn{5}{c}{$\nbh$} & \multicolumn{5}{c}{$\mbh\ [\msun]$} \vspace{0.07cm} \\
   & $-1\sigma$ & $-2\sigma$ & Mode & $+1\sigma$ & $+2\sigma$ & $-1\sigma$ & $-2\sigma$ & Mode & $+1\sigma$ & $+2\sigma$}
\startdata
NGC 0104 (47Tuc) &   0 &   8 & \textbf{ 55} & 137 & 242 &    0 &  130 & \textbf{ 1170} &  3020 &  5550 \\
NGC 0288         &   2 &  11 & \textbf{ 20} &  28 &  39 &   34 &  228 & \textbf{  444} &   664 &   909 \\
NGC 1261         &   5 &  27 & \textbf{ 53} &  81 & 109 &   61 &  560 & \textbf{ 1139} &  1735 &  2347 \\
NGC 1851         &   0 &   4 & \textbf{ 24} &  57 & 103 &    0 &   70 & \textbf{  510} &  1262 &  2290 \\
NGC 2298         &   0 &   1 & \textbf{  5} &  10 &  16 &    0 &   19 & \textbf{  101} &   229 &   395 \\
NGC 2808         &   0 &  36 & \textbf{114} & 201 & 294 &    0 &  546 & \textbf{ 2067} &  3978 &  6152 \\
NGC 3201         &   0 &   7 & \textbf{ 45} &  89 & 205 &    0 &  109 & \textbf{  981} &  2031 &  5322 \\
NGC 4147         &   0 &   1 & \textbf{  4} &   8 &  14 &    0 &   14 & \textbf{   83} &   206 &   358 \\
NGC 4590 (M68)   &   0 &   8 & \textbf{ 21} &  34 &  47 &    0 &  128 & \textbf{  418} &   742 &  1097 \\
NGC 4833         &   0 &  29 & \textbf{ 80} & 135 & 268 &    0 &  520 & \textbf{ 1734} &  3050 &  6625 \\
NGC 5024 (M53)   &   0 &  23 & \textbf{ 70} & 120 & 175 &    0 &  406 & \textbf{ 1443} &  2683 &  4100 \\
NGC 5053         &  21 &  30 & \textbf{ 84} & 106 & 159 &  451 &  657 & \textbf{ 2113} &  2640 &  3969 \\
NGC 5272 (M3)    &   0 &   6 & \textbf{ 39} &  98 & 173 &    0 &  110 & \textbf{  909} &  2379 &  4313 \\
NGC 5286         &   0 &   2 & \textbf{ 27} &  72 & 134 &    0 &   54 & \textbf{  659} &  1780 &  3334 \\
NGC 5466         &  11 &  24 & \textbf{ 43} &  90 & 156 &  162 &  457 & \textbf{  984} &  2121 &  4045 \\
NGC 5904 (M5)    &   0 &  22 & \textbf{ 70} & 126 & 191 &    0 &  383 & \textbf{ 1390} &  2648 &  4141 \\
NGC 5927         &  16 &  44 & \textbf{ 79} & 111 & 176 &  242 &  850 & \textbf{ 1610} &  2367 &  3935 \\
NGC 5986         &   3 &  59 & \textbf{114} & 172 & 268 &    0 & 1157 & \textbf{ 2489} &  3877 &  6183 \\
NGC 6093 (M80)   &   0 &   3 & \textbf{ 20} &  52 &  92 &    0 &   60 & \textbf{  479} &  1250 &  2251 \\
NGC 6101         &  60 &  83 & \textbf{100} & 152 & 190 & 1404 & 2005 & \textbf{ 2450} &  3996 &  4723 \\
NGC 6144         &   2 &  15 & \textbf{ 28} &  41 &  74 &    0 &  298 & \textbf{  619} &   948 &  1775 \\
NGC 6171 (M107)  &   0 &  13 & \textbf{ 32} &  60 & 107 &    0 &  250 & \textbf{  713} &  1326 &  2514 \\
NGC 6205 (M13)   &   0 &  61 & \textbf{127} & 194 & 343 &    0 & 1170 & \textbf{ 2768} &  4428 &  8388 \\
NGC 6218 (M12)   &   0 &  18 & \textbf{ 36} &  59 & 107 &    0 &  387 & \textbf{  847} &  1368 &  2508 \\
NGC 6254 (M10)   &   0 &  11 & \textbf{ 27} &  44 &  63 &    0 &  188 & \textbf{  568} &   981 &  1472 \\
NGC 6304         &   0 &   8 & \textbf{ 36} &  53 &  78 &    0 &  116 & \textbf{  372} &   646 &   930 \\
NGC 6341 (M92)   &   0 &   4 & \textbf{ 23} &  54 &  92 &    0 &   86 & \textbf{  530} &  1323 &  2303 \\
NGC 6352         &   0 &   4 & \textbf{ 10} &  17 &  27 &    0 &   69 & \textbf{  211} &   363 &   618 \\
NGC 6366         &   0 &   1 & \textbf{  4} &   8 &  15 &    0 &   21 & \textbf{   90} &   174 &   347 \\
NGC 6397         &   0 &   0 & \textbf{  2} &   7 &  14 &    0 &    0 & \textbf{   63} &   178 &   367 \\
NGC 6535         &   0 &   0 & \textbf{  1} &   2 &   4 &    0 &    1 & \textbf{   17} &    45 &    85 \\
NGC 6541         &   0 &   5 & \textbf{ 29} &  70 & 121 &    0 &  101 & \textbf{  679} &  1713 &  3018 \\
NGC 6584         &   0 &   7 & \textbf{ 25} &  44 &  65 &    0 &  137 & \textbf{  520} &  1002 &  1544 \\
NGC 6624         &   0 &   0 & \textbf{  1} &   3 &   6 &    0 &    0 & \textbf{   14} &    46 &   101 \\
NGC 6637 (M69)   &   0 &  25 & \textbf{ 57} &  82 & 120 &    0 &  466 & \textbf{ 1125} &  1689 &  2660 \\
NGC 6652         &   0 &   1 & \textbf{  4} &  10 &  18 &    0 &   11 & \textbf{   88} &   230 &   414 \\
NGC 6656 (M22)   &   0 &  10 & \textbf{ 57} & 114 & 325 &    0 &  168 & \textbf{ 1303} &  2683 &  8411 \\
NGC 6681 (M70)   &   0 &   4 & \textbf{ 14} &  28 &  46 &    0 &   70 & \textbf{  310} &   646 &  1087 \\
NGC 6715 (M54)   &   0 &   7 & \textbf{ 80} & 213 & 396 &    0 &  151 & \textbf{ 1966} &  5258 &  9862 \\
NGC 6717 (Pal9)  &   0 &   0 & \textbf{  1} &   4 &   7 &    0 &    2 & \textbf{   33} &    90 &   171 \\
NGC 6723         &   2 &  36 & \textbf{ 88} & 136 & 279 &    0 &  726 & \textbf{ 1837} &  3009 &  6691 \\
NGC 6752         &   0 &   0 & \textbf{  9} &  24 &  48 &    0 &    6 & \textbf{  226} &   627 &  1230 \\
NGC 6779 (M56)   &   1 &  13 & \textbf{ 28} &  44 &  57 &    0 &  240 & \textbf{  597} &   958 &  1302 \\
NGC 6809 (M55)   &   6 &  28 & \textbf{ 67} & 152 & 265 &    0 &  450 & \textbf{ 1498} &  3542 &  6805 \\
NGC 6838 (M71)   &   1 &   4 & \textbf{ 10} &  19 &  37 &    0 &   73 & \textbf{  222} &   420 &   884 \\
NGC 6934         &   0 &   4 & \textbf{ 16} &  31 &  49 &    0 &   77 & \textbf{  324} &   675 &  1077 \\
NGC 6981 (M72)   &  10 &  30 & \textbf{ 48} &  78 & 108 &  170 &  592 & \textbf{ 1017} &  1650 &  3100 \\
NGC 7078 (M15)   &   0 &   4 & \textbf{ 41} & 109 & 201 &    0 &   97 & \textbf{ 1022} &  2725 &  5028 \\
NGC 7089 (M2)    &   0 &   3 & \textbf{ 36} &  95 & 175 &    0 &   70 & \textbf{  868} &  2338 &  4368 \\
NGC 7099 (M30)   &   0 &   0 & \textbf{  6} &  17 &  34 &    0 &    2 & \textbf{  160} &   438 &   875 \\
\enddata
\tablecomments{Mode and mode-centric confidence intervals ($1\sigma$,$2\sigma$) are presented for $\nbh$ and $\mbh$ in each GC, using the Harris masses (computed from integrated V-band luminosities) in column 5 of \autoref{T:raw_results_dr} to convert from $\nbh/\ncluster$ and $\mbh/\mcluster$. These predictions are based on the mass segregation parameter $\Deltarfifty$.}
\label{T:bh_dr_Harris}
\end{deluxetable*}

\floattable
\begin{deluxetable*}{l|rrrrr|rrrrr}
\tabletypesize{\scriptsize}
\tablecolumns{11}
\tablewidth{0pt}
\tablecaption{Predicted Number and Mass of Retained BHs ($\Deltaa$+Harris)}
\tablehead{ & & & & & & & & & & \vspace{-4pt}\\
   \multirow{2}{*}{Cluster} & \multicolumn{5}{c}{$\nbh$} & \multicolumn{5}{c}{$\mbh\ [\msun]$} \vspace{0.07cm} \\
   & $-1\sigma$ & $-2\sigma$ & Mode & $+1\sigma$ & $+2\sigma$ & $-1\sigma$ & $-2\sigma$ & Mode & $+1\sigma$ & $+2\sigma$}
\startdata
NGC 0104 (47Tuc) &   0 &   8 & \textbf{ 53} & 136 & 238 &    0 &  160 & \textbf{ 1210} &  3130 &  5580 \\
NGC 0288         &   4 &  17 & \textbf{ 32} &  46 &  81 &   38 &  354 & \textbf{  691} &  1042 &  1924 \\
NGC 1261         &   6 &  27 & \textbf{ 52} &  81 & 106 &   90 &  569 & \textbf{ 1141} &  1719 &  2259 \\
NGC 1851         &   0 &   4 & \textbf{ 23} &  56 &  97 &    0 &   70 & \textbf{  492} &  1233 &  2165 \\
NGC 2298         &   0 &   2 & \textbf{  6} &  11 &  17 &    0 &   30 & \textbf{  123} &   247 &   398 \\
NGC 2808         &   0 &  25 & \textbf{ 92} & 173 & 261 &    0 &  380 & \textbf{ 1648} &  3491 &  5636 \\
NGC 3201         &   0 &   6 & \textbf{ 37} &  72 & 179 &    0 &   86 & \textbf{  805} &  1646 &  4580 \\
NGC 4147         &   0 &   0 & \textbf{  3} &   8 &  14 &    0 &   10 & \textbf{   76} &   196 &   350 \\
NGC 4590 (M68)   &   0 &  11 & \textbf{ 24} &  38 &  51 &    0 &  181 & \textbf{  500} &   831 &  1173 \\
NGC 4833         &   0 &  28 & \textbf{ 75} & 125 & 244 &    0 &  523 & \textbf{ 1633} &  2815 &  5909 \\
NGC 5024 (M53)   &   0 &  15 & \textbf{ 53} & 101 & 156 &    0 &  240 & \textbf{ 1068} &  2287 &  3720 \\
NGC 5053         &  13 &  22 & \textbf{ 39} &  90 & 145 &  252 &  430 & \textbf{  897} &  2202 &  3699 \\
NGC 5272 (M3)    &   0 &   5 & \textbf{ 38} &  98 & 174 &    0 &  104 & \textbf{  903} &  2391 &  4368 \\
NGC 5286         &   0 &   0 & \textbf{ 22} &  59 & 117 &    0 &    5 & \textbf{  536} &  1474 &  2937 \\
NGC 5466         &  20 &  32 & \textbf{ 88} & 115 & 176 &  424 &  647 & \textbf{ 1182} &  2824 &  4517 \\
NGC 5904 (M5)    &   0 &  15 & \textbf{ 55} & 106 & 166 &    0 &  246 & \textbf{ 1047} &  2214 &  3581 \\
NGC 5927         &  12 &  44 & \textbf{ 75} & 102 & 150 &  178 &  855 & \textbf{ 1509} &  2150 &  3299 \\
NGC 5986         &   8 &  62 & \textbf{115} & 170 & 240 &   16 & 1259 & \textbf{ 2533} &  3792 &  5335 \\
NGC 6093 (M80)   &   0 &   2 & \textbf{ 17} &  46 &  84 &    0 &   37 & \textbf{  415} &  1116 &  2070 \\
NGC 6101         &  73 &  91 & \textbf{104} & 156 & 193 & 1698 & 2255 & \textbf{ 2568} &  4124 &  4787 \\
NGC 6144         &   2 &  15 & \textbf{ 25} &  36 &  49 &   28 &  336 & \textbf{  594} &   854 &  1178 \\
NGC 6171 (M107)  &   3 &  17 & \textbf{ 42} &  66 & 111 &   16 &  342 & \textbf{  872} &  1446 &  2548 \\
NGC 6205 (M13)   &   0 &  59 & \textbf{119} & 179 & 294 &    0 & 1166 & \textbf{ 2597} &  4055 &  6989 \\
NGC 6218 (M12)   &   0 &  19 & \textbf{ 37} &  58 & 103 &    0 &  412 & \textbf{  840} &  1313 &  2193 \\
NGC 6254 (M10)   &   0 &  10 & \textbf{ 26} &  42 &  59 &    0 &  180 & \textbf{  544} &   939 &  1389 \\
NGC 6304         &   0 &   9 & \textbf{ 35} &  49 &  68 &    0 &  118 & \textbf{  361} &   599 &   839 \\
NGC 6341 (M92)   &   0 &   3 & \textbf{ 21} &  52 &  90 &    0 &   66 & \textbf{  490} &  1247 &  2240 \\
NGC 6352         &   0 &   4 & \textbf{ 10} &  18 &  28 &    0 &   68 & \textbf{  210} &   373 &   612 \\
NGC 6366         &   0 &   1 & \textbf{  3} &   6 &  11 &    0 &   13 & \textbf{   64} &   133 &   265 \\
NGC 6397         &   0 &   0 & \textbf{  1} &   3 &   6 &    0 &    0 & \textbf{   26} &    78 &   175 \\
NGC 6535         &   0 &   1 & \textbf{  2} &   3 &   4 &    0 &   10 & \textbf{   36} &    65 &    97 \\
NGC 6541         &   0 &   4 & \textbf{ 27} &  68 & 118 &    0 &   92 & \textbf{  653} &  1664 &  2965 \\
NGC 6584         &   0 &   4 & \textbf{ 17} &  36 &  58 &    0 &   67 & \textbf{  355} &   830 &  1408 \\
NGC 6624         &   2 &  66 & \textbf{ 78} &  91 & 105 &  123 &  326 & \textbf{  515} &   703 &   887 \\
NGC 6637 (M69)   &   1 &  21 & \textbf{ 46} &  67 &  89 &    0 &  400 & \textbf{  920} &  1375 &  1854 \\
NGC 6652         &   0 &   1 & \textbf{  4} &  11 &  18 &    0 &   13 & \textbf{   91} &   230 &   403 \\
NGC 6656 (M22)   &   0 &   9 & \textbf{ 61} & 125 & 352 &    0 &  125 & \textbf{ 1393} &  2903 &  9047 \\
NGC 6681 (M70)   &   0 &   3 & \textbf{ 12} &  24 &  39 &    0 &   57 & \textbf{  261} &   564 &   932 \\
NGC 6715 (M54)   &   0 &   0 & \textbf{ 48} & 133 & 281 &    0 &    0 & \textbf{ 1193} &  3326 &  7056 \\
NGC 6717 (Pal9)  &   0 &   0 & \textbf{  1} &   4 &   7 &    0 &    2 & \textbf{   33} &    87 &   166 \\
NGC 6723         &   8 &  44 & \textbf{ 92} & 136 & 276 &   79 &  886 & \textbf{ 1921} &  3014 &  6619 \\
NGC 6752         &   0 &   0 & \textbf{  8} &  23 &  45 &    0 &    2 & \textbf{  211} &   584 &  1156 \\
NGC 6779 (M56)   &   2 &  15 & \textbf{ 31} &  46 &  61 &    8 &  276 & \textbf{  645} &  1014 &  1358 \\
NGC 6809 (M55)   &   7 &  31 & \textbf{ 68} & 158 & 269 &   29 &  517 & \textbf{ 1523} &  3744 &  6891 \\
NGC 6838 (M71)   &   6 &  10 & \textbf{ 14} &  27 &  42 &  110 &  199 & \textbf{  297} &   611 &   995 \\
NGC 6934         &   0 &   4 & \textbf{ 17} &  33 &  51 &    0 &   88 & \textbf{  349} &   706 &  1110 \\
NGC 6981 (M72)   &  13 &  29 & \textbf{ 50} &  70 & 102 &  263 &  613 & \textbf{ 1072} &  1555 &  2355 \\
NGC 7078 (M15)   &   0 &   0 & \textbf{ 30} &  81 & 162 &    0 &    0 & \textbf{  746} &  2044 &  4112 \\
NGC 7089 (M2)    &   0 &   0 & \textbf{ 27} &  74 & 148 &    0 &    0 & \textbf{  623} &  1736 &  3514 \\
NGC 7099 (M30)   &   0 &   0 & \textbf{  5} &  15 &  31 &    0 &    0 & \textbf{  137} &   383 &   784 \\
\enddata
\tablecomments{Mode and mode-centric confidence intervals ($1\sigma$,$2\sigma$) are presented for $\nbh$ and $\mbh$ in each GC, using the Harris masses (computed from integrated V-band luminosity) in column 5 of \autoref{T:raw_results_dr} to convert from $\nbh/\ncluster$ and $\mbh/\mcluster$. These predictions are based on the mass segregation parameter $\Deltaa$.}
\label{T:bh_dA_Harris}
\end{deluxetable*}

\end{document}